\documentclass[aps,pre,preprint,groupedaddress,showpacs]{revtex4-1}
\usepackage[utf8]{inputenc}
\usepackage[T1]{fontenc}

\usepackage{amsmath}
\usepackage{amssymb}
\usepackage{graphicx}
\usepackage{verbatim}

\begin{document}

\title{Effective one-component model of binary mixture: molecular arrest induced by the spatially correlated stochastic dynamics}


\author{M. Majka}
\email{maciej.majka@uj.edu.pl}
\affiliation{Marian Smoluchowski Institute of Physics, Jagiellonian University, ul. prof. Stanis\l{}awa \L{}ojasiewicza 11, 30-348 Krak\'{o}w Poland}
\author{P. F. G\'{o}ra}
\affiliation{Marian Smoluchowski Institute of Physics, Jagiellonian University, ul. prof. Stanis\l{}awa \L{}ojasiewicza 11, 30-348 Krak\'{o}w Poland}


\begin{abstract}
Spatially correlated noise (SCN), i.e. the thermal noise that affects neighbouring particles in a similar manner, is ubiquitous in soft matter systems. In this work, we apply the over-damped SCN-driven Langevin equations as an effective, one-component model of the dynamics in dense binary mixtures. We derive the thermodynamically consistent fluctuation-dissipation relation for SCN to show that it predicts the molecular arrest resembling the glass transition, i.e. the critical slow-down of dynamics in the disordered phases. We show that the mechanism of singular dissipation is embedded in the dissipation matrix, accompanying SCN. We are also able to identify the characteristic length of collective dissipation, which diverges at critical packing. This novel physical quantity conveniently describes the difference between the ergodic and non-ergodic dynamics. The model is fully analytically solvable, one-dimensional and admits arbitrary interactions between the particles. It qualitatively reproduces several different modes of arrested disorder encountered in binary mixtures, including e.g. the re-entrant arrest. The model can be effectively compared to the mode coupling theory.
\end{abstract} 

\maketitle

\section*{Introduction}

While some phase transitions in soft matter systems, e.g. crystallization or demixing, are sufficiently explained by the minimizing of free energy \cite{bib:rev1, bib:rev2,bib:likos}, the glass transition\cite{bib:rev1, bib:rev2} seems to stem from changes in the microscopic dynamics. The works of Mori and Zwanzig have shown that the diffusion of particles in densely packed systems might involve both memory and strong collectivity \cite{bib:MZ1,bib:MZ2}. While the temporal aspect is well recognized in research on anomalous diffusion \cite{bib:SCKou}, the spatio-collective aspect has drawn attention mostly in the context of hydrodynamics\cite{bib:Dhont, bib:Deutch,bib:hydrodynamic}. However, the collective effects are not restricted to continuous media. The collectivity in diffusion means that thermal fluctuations affecting the nearby particles take the form of Spatially Correlated Noise (SCN) \cite{bib:SCN1,bib:SCN2,bib:polymers}. The sources of SCN include e.g. hydrodynamic interactions\cite{bib:Deutch,bib:hydrodynamic,bib:hs1,bib:hs2}, stirring by active particles \cite{bib:cell1,bib:cell2}, local density fluctuations \cite{bib:miyazaki} and phase transitions. Recently, the current authors have also shown that SCN might arise as the dynamic counterpart of entropic interactions\cite{bib:SCN1,bib:SCN2} in binary mixtures (also named excluded volume or effective interactions\cite{bib:majka, bib:likos, bib:lekkerkerker}). In this work, we build upon these results and investigate the SCN-driven Langevin equations as the possible one-component model for the effective dynamics in binary mixtures. We will show that this model predicts the effects of arrested disorder, showing that SCN can be efficiently employed to model the complex behaviour of these systems.

Let us begin by recalling the general phenomenology of colloidal glasses, as their general properties apply to binary mixtures as well. Glasses are disordered systems with extremely slow molecular dynamics \cite{bib:rev1, bib:rev2}. As the packing fraction of particles increases above the critical value for crystallization (0.494 for the mono-disperse hard spheres \cite{bib:rev1}), the system enters the supercooled state in which its viscosity grows by 3-4 orders of magnitude\cite{bib:rev1, bib:rev2, bib:pusey, bib:phi,bib:likos_diff}. Eventually, at a packing fraction of $\simeq 0.58-0.62$, the system becomes a glass, the relaxation times become virtually infinite and equilibration is no longer possible\cite{bib:mass_sims}. The exact critical values for this transition depends on the type of the system, but also on the experimental protocol \cite{bib:ritland,bib:bartenev}. Thus, glasses are seen as an inherently out-of-equilibrium state of matter\cite{bib:non_eq}. Remarkably, the polydispersity of particles significantly facilitates the process of vitrification\cite{bib:rev2}. Glasses are also characterized by dynamic heterogeneity, i.e. the coexistence of domains with significantly different mobility \cite{bib:rev1, bib:rev2}. This manifests in intermittent diffusion \cite{bib:subdiff1,bib:subdiff2}, the correlated displacement of particles \cite{bib:doliwa, bib:weeks, bib:dim}, the presence of mobility clusters \cite{bib:cluster} and cooperative rearrangement \cite{bib:coop}. The dynamics within these domains is very strongly spatially correlated, usually involving $\simeq$10-20 neighbouring particles \cite{bib:weeks,bib:donth,bib:nature}, in contrast to only $\simeq$6 in the liquid phase. There are numerous theories approaching the glass transition, but the main insights have been provided by mode coupling theory (MCT) \cite{bib:rev1,bib:mct, bib:rfot} and the theory of multi-point correlation functions \cite{bib:4point}. The former is especially successful in predicting the two-step relaxation in the dynamic density correlation function \cite{bib:mct,bib:relax,bib:aging} from first principles. However, exactly solvable models of the glass transition are very scarce \cite{bib:spin, bib:hs1, bib:hs2, bib:leutheusser} and a few fundamental questions still lack a definite answer. One open issue is the existence and nature of the order parameter for this transition\cite{bib:rev1}. Another problem is the existence of a divergent length-scale accompanying the transition, i.e. while it is postulated by MCT \cite{bib:rfot} and the affine transformation theory \cite{bib:mosayebi}, it has not been observed directly and its necessity is also contested \cite{bib:wyart}. Finally, the relation between the dynamic heterogeneity and the high viscosity is also not fully understood.

Binary mixtures introduce several further complications into this picture. These are systems composed of two different species of particles, usually suspended in a fluid. They exhibit a remarkable variety of thermodynamic phases. These include mixing and demixing effects \cite{bib:majka,bib:majka2} in the double fluid phase, crystallization \cite{bib:binmix_cryst} as well as a few types of the arrested disordered \cite{bib:binmix2}: the single glass (formed by the large particles at a high volume fraction, with the smaller particles caged in-between), double glass (a cooperative form of arrest involving both large and small particles) and asymmetric glass (large particles are sparse and embedded in a glass of smaller particles). Binary mixtures are also well known for their reentrant transition, i.e. while the total packing fraction is kept constant, the partial replacement of the bigger particles by the smaller ones turns the vitrified system into a fluid and again into a glass as the total packing fraction is increased. Binary mixtures are also the minimal, controllable setting for the investigation of how polydispersity affects the glass transition. The glassy behaviour of binary mixtures is best documented for three- and two-dimensional systems\cite{bib:binmix2,bib:binmix1,bib:binmix3,bib:binmix2D} and to a lesser extent in the one-dimensional case (both classical\cite{bib:binmix_c} and quantum\cite{bib:binmix_q}). 

While some theoretical insight into the arrest of binary mixtures has been obtained via MCT and simulations\cite{bib:binmix1}, our one-component SCN-based model aims at understanding the effective dynamics at the level of individual particles. We will identify the collective physical quantities that govern this dynamics. In the effective description the larger particles are chosen as observed, while the smaller ones (and solvent, if present) are treated as the thermal bath. Thus, their interactions with the observed particles are replaced by the correlated thermal noise, dissipation and effective modification to the deterministic forces. What follows is that, in effective theory, the arrest of bigger particles must emerge from the properties of the thermal noise and friction. Indeed, in this work we will show how this can happen. We derive a new variant of the fluctuation-dissipation relation (FDR) of the second kind (i.e. the interdependence between friction and noise correlation function\cite{bib:kubo}) for the SCN-driven, over-damped Langevin equations. SCN entails a collective form of friction, i.e. the friction matrix, which describes how the larger particles affect each other's dissipation via a common complex medium. We will show that in the disordered phases this FDR predicts a huge rise in viscosity, by 3-4 orders of magnitude, when some critical combinations of density and noise correlation length are met. This mechanism is universal, i.e. we establish it for an arbitrary choice of interaction potentials and noise correlation function. The SCN-driven Langevin dynamics investigated in this paper are technically similar to the Stokesian dynamics, which describe the mono-disperse colloidal particles suspended in a continuous solvent\cite{bib:Dhont,bib:Deutch,bib:hydrodynamic}. While the FDR for this case is already known\cite{bib:Deutch}, it does not predict arrest in the highly packed systems. Conversely, our model addresses the mixture of particles which are comparable in size, with a ratio of diameters possibly equal to e.g. $0.9-0.01$. The effective thermal bath composed of such particles is dominated by the excluded volume interactions hence it does not act like a purely continuous medium. Thus a different FDR is necessary to describe this regime.

Our model leads to several significant results. We devote most space to the analysis of the SCN-induced transition, embedded in the FDR, as it provides a novel perspective on the possible mechanism of the arrest in disordered systems. While MCT attributes the arrest to the memory effects and the correlated temporal evolution in the system, our model suggests that the spatial aspect might be equally important. The fact that pure, non-correlated in time SCN can induce arrest has not been recognized before. We will show that the SCN-induced arrest reproduces the transition between ergodic and non-ergodic dynamics, resembling the transition from the supercooled fluid into the glassy phase. It also reveals, what we name, the collective dissipation length. This magnitude describes the typical length-scale within which the larger particles affect each others' dissipation. We will show that it becomes divergent as the system enters the arrested state, even though e.g. the noise correlation length remains finite. As this quantity is related to the dynamics, not to the structure, it introduces a novel perspective into the problem of the divergent length-scale accompanying the glass transition. The properties of SCN-induced arrest will be illustrated with two exemplary systems: soft particles and hard spheres. Further, we will also identify two additional, auxiliary arrest mechanisms embedded in our SCN-driven model. What follows, this model is able to reproduce the major part of the binary mixture phase-diagram. The comparison with experiment is only qualitative, as the current model is one-dimensional, but the analogues of the main disordered phases can be conveniently recognized, including the re-entrant transition. Eventually, the SCN-induced arrest is discussed in the context of glass research. This leads to the conclusion that our model is related to MCT, but it utilizes a significantly different set of approximations. The SCN-induced arrest is identified as a simplified variant of the ideal glass transition, described by MCT. While, from the modern perspective, this makes our model insufficient as a stand-alone theory of the physical glass transition, the explicit form of dynamics obtained in this work stands as a possibly useful tool in the modelling of the dense binary mixtures. Its role is similar to the role of Stokesian dynamics in mono-disperse colloids. On a general level, our results show that phase transitions can be efficiently embedded in the FDR, thus it is possible to describe a complex, multicomponent system with a simpler, less computationally expansive model, without losing its vital characteristics.

 The paper is organized as follows: in the next two sections \ref{sec:derivation} the theory is derived and the transition mechanism is analysed in a general way; \ref{sec:example} further our results are illustrated with the behaviour of two systems: soft particles and hard spheres. In the following section, the modes of binary mixture dynamics covered by our model are addressed and compared to experimental data. In the final sections \ref{sec:MCT}, the relationship with MCT is discussed and, eventually, we place our model in the broader context of glass research. Appendices \ref{app:stochastic}-\ref{app:mct} contain the detailed calculations used throughout the derivations.

\section*{Fluctuation-dissipation relation (FDR) for SCN}\label{sec:derivation}
As the first step, we will introduce the approximation of SCN-driven dynamics in the context of binary mixtures. We consider a set of $N$ observed particles of mass $\mathcal{M}$, at positions $x_i$, in contact with the thermal bath of $N'$ particles with mass $\mathcal{M}'$ at positions $x'_i$. The full, microscopic dynamics of this system reads:
\begin{align}
\mathcal{M}\ddot x_i=-\sum_j^N \partial_{x_i}U_0(x_i-x_j)-\sum_j^{N'}\partial_{x_i}V(x_i-{x'}_j)&&\mathcal{M}' \ddot {x'}_i=-\sum_j^{N'}\partial_{{x'}_i}v({x'}_i-{x'}_j)-\sum_j^N\partial_{{x'}_i}V(x_i-{x'}_i)
\end{align}
where $U_0(r)$ is the direct microscopic interaction of the larger particles, $V(r)$ is the coupling force and $v(r)$ is the internal interaction of the thermal bath. We will use $r$ as a placeholder for relative distance throughout the paper. Using the projection operator method, the dynamics of the observed particles can be formally reduced to a set of Generalized Langevin Equations\cite{bib:MZ1,bib:MZ2,bib:Deutch,bib:SCN2}, which have no explicit dependence on $x'_i$, i.e.:
\begin{align}
\mathcal{M} \ddot x_i+\sum_j^N \int_{-\infty}^t dt' G(x_i-x_j,t-t')\dot x_j(s)=\sum_j^NF(x_i-x_j)+\xi_i(t) \label{eq:exactGLE}&&<\xi_i(t)\xi_j(t')>=\beta^{-1}G(x_i-x_j,t-t') 
\end{align}
Here, $\beta^{-1}=k_B T$, where $k_B$ is the Boltzmann constant, $T$ is temperature and $<...>$ denotes the average. The formal definitions of $F(r)$, $G(r,t-t')$ and $\xi_i(t)$ as functions of $V(r)$, $v(r)$ and other microscopic parameters can be found in many standard textbooks nowadays\cite{bib:MZ1,bib:MZ2,bib:Deutch,bib:SCN2}. However, these exact expressions are rarely applied in practice as they require full microscopic information about the system, which is usually unavailable. Instead, these terms have a well-established general interpretation that can be employed to formulate the effective dynamics. $\xi_i(t)$ acts as the thermal noise (i.e. a fluctuating external force), $G(r,t-t')$ is the collective dissipation memory kernel and the relation between $<\xi_i(t)\xi_j(t')>$ and $G(x_i-x_j,t-t')$ provided in \eqref{eq:exactGLE} is the FDR, indicating that $\xi_i(t)$ is spatio-temporally correlated. $F(r)$ can be decomposed into a combination of microscopic and effective forces, i.e. $F(r)=F_0(r)+F_{eff}(r)$, where $F_0(r)=-\partial_rU_0(r)$ and $F_{eff}(r)=-\partial_rU_{eff}$. In equilibrium, $U_{eff}$ can be calculated from the partition function\cite{bib:majka}:
\begin{align}
U_{eff}=-\beta^{-1}\ln \Xi&& \Xi=\int dx'_1...dx'_N \exp \left[ -\beta \left(\sum_i^N\sum_j^{N'}V(x_i-x'_j)+\sum_{i>j}^{N'}v(x'_i-x'_j)\right)\right] \label{eq:Ueff}
\end{align}
which is also a formal definition of the effective potential. In general, $U_{eff}$ might contain also the three- and greater body interactions\cite{bib:likos}, but we will restrict our considerations to the pair potentials. We will denote $U(r)=U_0(r)+U_{eff}$, so $F(r)=-\partial_rU(r)$.
 
The dynamics in \eqref{eq:exactGLE} have been a subject to various simplifying approximations. Most importantly, $\xi_i(t)$ can be replaced by a correlated Gaussian noise, which turns \eqref{eq:exactGLE} into stochastic differential equations. Further, Deutch and Oppenheim established an approximation that neglects the memory effects, while maintaining the collective dissipation\cite{bib:Deutch}: 
\begin{equation}
\mathcal{M} \ddot x_i(t)+\sum_j^N \Gamma(x_i-x_j,t) \dot x_j(t)=\sum_j^N F(x_i-x_j)+\xi_i(t) \label{eq:Deutch}
\end{equation}
This approximation is valid when the colloidal particles are much heavier than the thermal bath particles and at time scales larger than $\bar r/v_s$, where $v_s$ is the speed of sound in the thermal bath and $\bar r$ is the average distance between the colloidal particles \cite{bib:Deutch}. Under these conditions, $\xi_i(t)$ effectively behaves as SCN. Deutch and Oppenheim also showed that the FDR for equation \eqref{eq:Deutch} is slightly different from the standard one (see e.g. equation 4.13 in their work\cite{bib:Deutch}). Eventually, they employed the theory of hydrodynamic interactions (i.e. the interaction of a particle with the velocity field of a fluid) to establish the explicit form of the FDR for a colloidal particle in a continuous medium. 

However, the dynamics in \eqref{eq:Deutch} are not restricted to the hydrodynamic regime and to continuous media. In binary mixtures of particles with comparable sizes, the effective interactions, \eqref{eq:Ueff}, may play a dominant role. In our previous works, we have shown that the microscopic coupling force between an observed particle and the thermal bath, $F_{c}(x_i)=-\sum_{j}^{N'}\partial_{x_i}V(x_i-x'_j)$, is correlated like effective forces, i.e. $<F_c(x_i)F_c(x_j)>=<\sum_{n,n'}^NF_{eff}(x_i-x_n)F_{eff}(x_j-x_{n'})>$. We also postulated that this correlation function could be adopted for SCN, as $\xi_i(t)$ is supposed to approximate $F_c(x_i)$, thus:
\begin{equation}
<\xi_i(t)\xi_j(t')>\simeq<\sum_{n,n'}^NF_{eff}(x_i-x_n)F_{eff}(x_j-x_{n'})>=\sigma^2\delta(t-t') H(x_i-x_j) \label{eq:H}
\end{equation}
where $\sigma$ is the noise amplitude and $H(r)$ is the dimensionless correlation function. This dependence links the microscopic picture to the effective description, i.e. the microscopic potentials define the effective forces, which, in turn, define the correlation function for SCN. A particular form of $H(r)$ depends on the system under scrutiny, but, as a rule of thumb, one can expect that the range of correlations in noise is similar to the range of effective interactions. For example, in hard-sphere mixtures (Asakura-Oosawa model\cite{bib:lekkerkerker}) it is equal to the diameter of the smaller spheres, in the mixtures of polymers (Gaussian particle model) it is determined by the gyration radius of the longer chain\cite{bib:majka}, while in the ionic mixtures the range depends on the charges of particles and can exceed a few diameters of the larger particle\cite{bib:majka}. On the other hand, when the smaller particles are densely packed their motion becomes spatially correlated and this structural correlation function can be also adapted as an approximation for $<\xi_i(t)\xi_j(t')>$. Usually, it has an exponential profile with decay length ranging from 2-3 diameters of the smaller particle in the supercooled state up to 10 and higher at the transition point\cite{bib:doliwa,bib:weeks}. Nevertheless, in this work the derivation of the FDR is carried out for any choice of $H(r)$, thus our approach applies to a variety of systems. It should also be mentioned that for binary mixtures in a solvent, $\Gamma(r)$ can easily contain the Stokesian hydrodynamic friction $\gamma$, e.g. as a diagonal contribution. We will later show that the presence of solvent is embedded in our model and the standard relation $\beta \sigma^2/2=\gamma$ holds. This is achieved by tuning the model to reduce to ordinary Brownian diffusion in the limit of the non-correlated noise.

The correlation function in \eqref{eq:H} could be directly employed in the theory of Deutch and Oppenheim, but, as we aim to describe the arrest effect, we are interested in systems with a high packing fraction. Thus, in this work we focus our attention on the over-damped dynamics:
\begin{align}
\sum_j^N\Gamma(x_i-x_j)\dot x_j=\sum_{j\neq i}^N F(x_i-x_j)+\xi_i(t) \label{eq:Lan}
\end{align}
Functions $\Gamma(r)$ form an $N\times N$ friction matrix $\pmb{\Gamma}$, such that $\pmb{\Gamma}_{ij}=\Gamma(x_i-x_j)$. Similarly, from $H(r)$ one can form an $N\times N$ correlation matrix $\textbf{H}$, such that $\textbf{H}_{ij}=H(x_i-x_j)$. An important observation is that under, what we call, the constraint of thermodynamic consistency, the relation between $\Gamma(r)$ and $H(r)$ must become different from \eqref{eq:exactGLE} and the FDR established by Deutch and Oppenheim. Thermodynamic consistency means that if we demand the Fokker-Planck equation for \eqref{eq:Lan} to be satisfied by the Boltzmann distribution in the stationary (though possibly non-arrested) state, then we must redefine the FDR to achieve agreement. As we will show, the new FDR predicts the arrest effect in certain conditions. However, this strategy requires a comment. The need to modify the FDR comes from the fact that the over-damped dynamics is qualitatively different from the inertial one. Our approach is a workaround for a formal transition between the inertial and non-inertial limit, which is a highly non-trivial problem for correlated noises\cite{bib:inertial_to_overdamped}. Another issue is that SCN can be represented as a linear combination of multiplicative noises, i.e. the sum of noise terms whose amplitude varies with $x_i$\cite{bib:SCN1,bib:SCN2}. For multiplicative noise one must also specify the \emph{interpretation} of the stochastic integral, i.e. how this integral mixes the past and future values of a stochastic process\cite{bib:gardiner}. Interpretation modifies the drift terms in the Langevin equations, so its choice should be supported by a relevant physical reasoning. In particular, it was realized during the study of the diffusion in viscosity landscapes\cite{bib:lau,bib:oded} (which also involves multiplicative noise and spatially variant dissipation) that the thermodynamic consistency can define the interpretation\cite{bib:lau}. In other words, one can assume a certain FDR \emph{a priori} and then adjust the noise interpretation (usually in some non-standard manner, i.e. neither Ito nor Stratonovich) to meet the consistency condition. Thus, in general, the FDR and interpretation act interchangeably in this context. For the diffusion in viscosity landscapes, the additional drifts preserve the equilibrium between the regions of different viscosity\cite{bib:lau}, thus their presence is physically justified. However, we will show that SCN has a peculiar property of not being sensitive to a change of interpretation. It behaves similarly to the additive noise\cite{bib:gardiner} and introduces no additional drifts. Thus, our strategy to determine the FDR from thermodynamic consistency is unambiguous for SCN. 

We will now proceed with a sketch derivation of our FDR, with full details provided in the appendixes \ref{app:stochastic}-\ref{app:FP}. As a preliminary step, let us assume that the system is one-dimensional, with size $L$ and density $N/L=\rho$. Using the function:
\begin{equation}
Q_{ik}=\frac{1}{\sqrt{N}}\left(\cos \frac{2\pi k x_i}{L}+\sin \frac{2\pi k x_i}{L}\right)
\end{equation}
we can approximate $\textbf{H}$ and $\pmb{\Gamma}$ via the Fourier series expansions, i.e.:
\begin{equation}
\begin{aligned}
\textbf{H}_{ij}\simeq\rho \sum_{k=-(M-1)/2}^{(M-1)/2}\hat H_k Q_{ik}Q_{jk}&&\hat H_k=\int_{-L/2}^{L/2}drH(r)\cos\frac{2\pi k r}{L} \\
\pmb{\Gamma}_{ij}\simeq\rho \sum_{k=-(M-1)/2}^{(M-1)/2}\hat \Gamma_k Q_{ik}Q_{jk}&&\hat \Gamma_k=\int_{-L/2}^{L/2}dr\Gamma(r)\cos\frac{2\pi k r}{L}
\end{aligned}
\end{equation}
where $M$ is a certain cut-off frequency. This cut-off is introduced as it is expected that the length-scales much finer than the size of the thermal bath particles should not play a role in the system. It also allows us to use a finite-dimensional matrix representations for $\textbf{H}$ and $\pmb{\Gamma}$, i.e.:
\begin{align}
\textbf{H}\simeq Q\Lambda_H Q^T && \pmb{\Gamma}\simeq Q\Lambda_\Gamma Q^T \label{eq:h_rep}
\end{align}
where $Q$ is the $N\times M$ rectangular matrix composed of $Q_{ik}$, while $\Lambda_H$ and $\Lambda_{\Gamma}$ are the $M\times M$ diagonal matrices with entries $\Lambda_{H,k}=\rho \hat H_k$ and $\Lambda_{\Gamma,k}=\rho \hat \Gamma_k$. Using the representation in \eqref{eq:h_rep}, we can generate the vector of the SCN, $\vec{\xi}$, from the vector of the non-correlated Gaussian noise $\vec{\eta}$, i.e.:
\begin{equation}
\vec{\xi}=\sigma Q \Lambda_{H}^{1/2} \vec{\eta} \label{eq:xi}
\end{equation}
 where $<\eta_i(t)\eta_j(t')>=\delta_{ij}\delta(t-t')$ and $\Lambda_{H}^{1/2}\Lambda_{H}^{1/2}=\Lambda_H$. One can calculate $<\vec{\xi}\vec{\xi}^T>$ to see that it equals $\sigma^2\delta(t-t')\textbf{H}$, but  since $Q_{ik}$ depends on $x_i$, this result is recovered only under the Ito interpretation, i.e. $x_i$ in a moment $t$ must be independent from the stochastic process increment $\vec{\eta}(t)$. 

Another idea, which is necessary for our derivation, is stochastic orthogonality. Matrix $Q$ is not orthogonal by itself, but in completely disordered systems, such that the distribution of $x_i$ is homogeneous and reads $p(x_i)\simeq L^{-1}$, it satisfies:
\begin{align}
\lim_{N\to +\infty} Q^TQ=\textbf{1}_M&&QQ^T=\textbf{1}_N \label{eq:Q_props}
\end{align}
The first equality is guaranteed by the central limit theorem, i.e. $[Q^TQ]_{kk'}=\sum_i^N Q_{ik}Q_{ik'}\to \delta_{kk'}$ as $N\to +\infty$ (see Appendix \ref{app:stochastic}). What follows, we also have $QQ^T\textbf{A}=QQ^TQ\Lambda_A Q^T=\textbf{A}$ hence $QQ^T$ acts as $\textbf{1}_N$, which is the other equality. With the aid of stochastic orthogonality we can invert $\pmb{\Gamma}$, i.e. $\pmb{\Gamma}^{-1}=Q\Lambda_\Gamma^{-1}Q^T$. However, it should be emphasized that stochastic orthogonality holds only in disordered systems. Thus, by utilizing this property in the calculations we automatically narrow our considerations to the disordered phases.

Inserting \eqref{eq:xi} in \eqref{eq:Lan}, multiplying by $\pmb{\Gamma}^{-1}$ and specifying the Ito interpretation, we transform the initial set of equations $\eqref{eq:Lan}$ into:
 \begin{equation}
 \dot{\vec{x}}=Q\Lambda_{\Gamma}^{-1}Q^T\vec{F}+\vec{C}+\sigma Q \Lambda^{-1}_{\Gamma} \Lambda_H^{1/2}\vec{\eta}  \label{eq:Lan1.5}
 \end{equation}
where $F_i=\sum_{j\neq i}^N F(x_i-x_j)$ and $\vec{C}$ is a correction term that might arise while switching from the initial, unknown, noise interpretation to the Ito interpretation. Since $Q_{ik}$ is the function of $x_i$, system \eqref{eq:Lan} has been turned into the usual system of stochastic differential equations with multiplicative noise. We will now specify $\vec{C}$, which is discussed in detail in Appendix \ref{app:stoch_int}. For a multivariate stochastic differential equation of the type $\dot x_i=f_i(\vec{x})+\sum_k g_{ik}(\vec{x})\eta_k$, the correction term reads\cite{bib:lau} $C_i=\alpha \sum_{j,k}g_{kj}(\vec{x})\partial_{x_j}g_{ik}(\vec{x})$, where $\alpha\in [0,1]$ defines the type of interpretation (i.e. $\alpha=0$ for Ito, $\alpha=1/2$ for Stratonovich). Applying this to \eqref{eq:Lan1.5}, we obtain the following result:
\begin{equation}
C_i=\alpha\sigma^2\sum_{k=-(M+1)/2}^{k=(M+1)/2} \frac{2\pi k}{L}Q_{ik}Q_{i,-k}\Lambda_{\Gamma,k}^{-2}\Lambda_{H,k}=0
\end{equation}
which is due to the antisymmetric nature of the expression under the sum. This means that the initial interpretation of the noise in \eqref{eq:Lan} is irrelevant, as  any choice leads to no additional drift. Thus, SCN behaves as additive noise. In this respect, models with SCN (where correlations depend on the inter-particle distances) are fundamentally different from models with a spatially variant diffusion coefficient\cite{bib:lau}. The noise interpretation cannot provide for thermodynamic consistency and it must be ensured by the appropriate FDR.

The next step is to write the stationary Fokker-Planck equation\cite{bib:Deutch,bib:gardiner} for \eqref{eq:Lan1.5}:
\begin{equation}
0=-\sum_i^N \partial_{x_i}\left( \sum_j^N \sum_{k=-(M-1)/2}^{(M-1)/2} Q_{ik} \Lambda_{\Gamma,k}^{-1} Q_{jk} F_j P_s \right)+\frac{\sigma^2}{2}\sum_{i,j}^N \partial_{x_i x_j}^2 \left(\sum_{k=-(M-1)/2}^{(M-1)/2} Q_{ik}\Lambda_{\Gamma,k}^{-1}\Lambda_{H,k}\Lambda_{k,\Gamma}^{-1}Q_{jk} P_s \right)  \label{eq:FPE0}
\end{equation}
The crucial assumption now is that the system can equilibrate, i.e. that $P_s$ is the Boltzmann distribution itself:
\begin{equation}
P_s=\mathcal{N}^{-1}\exp\left(-\frac{\beta}{2}\sum_{i,j}^NU(x_i-x_j) \right) \label{eq:Ps}
\end{equation}
where $\mathcal{N}$ is the normalization constant. When $P_s$ is known, \eqref{eq:FPE0} becomes the equation that defines the unknown spectrum $\Lambda_\Gamma$ and thus the FDR for SCN. The equilibration is possible in supercooled colloids, below the random close packing limit \cite{bib:rev1}, though it is extremely slow. The above method of defining the FDR is fully consistent when $H(r)$ is the equilibrium noise correlation function.

Obtaining $\Lambda_\Gamma$ from \eqref{eq:FPE0} is a highly technical task, which we discuss in detail in the Appendix \ref{app:FP}. Throughout this derivation, the non-correlated part of the dynamics (corresponding to the Stokesian friction and Brownian motion) is separated from the influence of spatial correlations, to show that the classical dissipation relation $\beta \sigma^2/2=\gamma$ still holds. This also means that the presence of hydrodynamic friction (i.e. solvent) is embedded in our model. Eventually, we obtain the main result of this paper, the FDR connecting matrices $\pmb{\Gamma}$ and $\textbf{H}$, which, in the thermodynamic limit $N\to +\infty$, $L\to+\infty$ (while $N/L=\rho$), reads:
\begin{equation}
\begin{split}
\pmb{\Gamma}_{ij}&=\Gamma(x_i-x_j)\simeq\frac{\gamma a_H}{\pi \rho}\int_0^{\frac{\pi m}{d}}dk \frac{\hat h(k)[2+\beta\rho \hat U(k)]\cos k(x_i-x_j)}{1+\hat h(k)+\beta\rho \hat U(k)}
\end{split}  \label{eq:Gamma_fin}
\end{equation}
Here $\hat h(k)=\rho \hat H(k)/a_H$, $\hat H(k),\hat U(k)$ are the Fourier transforms of $U(r)$ and $H(r)$, respectively, and $d$ is the particle diameter. $a_H$ is defined by the residual behaviour of $H(r)$ in the non-correlated limit, i.e. assuming that $H(r)$ depends on a certain noise correlation length $\lambda$, we have $\lim_{\lambda\to 0}H(r)=\frac{a_H}{\rho} \delta(r)$. The formula \eqref{eq:Gamma_fin} provides the elements of the friction-response matrix and defines the FDR that we are looking for.


\section*{The mechanism of SCN-induced arrest}\label{sec:mechanism}
There are several important properties of \eqref{eq:Gamma_fin} that we will now discuss. First of all, for $\hat h(k)=1$, which is the non-correlated case, \eqref{eq:Gamma_fin} reduces to $\pmb{\Gamma}_{ij}=\frac{\gamma a_H}{\rho}\delta(x_i-x_j)$, i.e. only its diagonal part is non-zero, as expected and the friction is of purely hydrodynamic origin. Further, in Appendix \ref{app:exponent}, we show that the mean behaviour of matrix $\pmb{\Gamma}$ is dominated by its diagonal terms. Finally, and most importantly, for the finite-range correlations, \eqref{eq:Gamma_fin} contains the mechanism of the singular dissipation. Let us introduce the joint temperature-packing parameter $\psi=\beta \rho$. Let us also denote the integrand of \eqref{eq:Gamma_fin} for the diagonal terms ($i=j$) as:
\begin{equation}
I(k,\psi)=\frac{ a_H}{\pi \rho}\frac{\hat h(k)[2+\psi \hat U(k)]}{1+\hat h(k)+\psi \hat U(k)} \label{eq:I}
\end{equation}
so $\pmb{\Gamma}_{ii}/\gamma=\int_0^{\frac{\pi m}{d}}dk I(k,\psi)$. $\pmb{\Gamma}_{ii}$ will develop the extremely high values provided that $I(k,\psi)$ has a non-removable singularity for a certain $k\in[0,\pi m/d]$. This happens when the denominator of $I(k,\psi)$ satisfies:
\begin{equation}
f(k,\psi)=1+\hat h(k)+\psi \hat U(k)=0 \label{eq:cond}
\end{equation} 

\begin{figure}[h]
\centering
\includegraphics[width=0.98\textwidth]{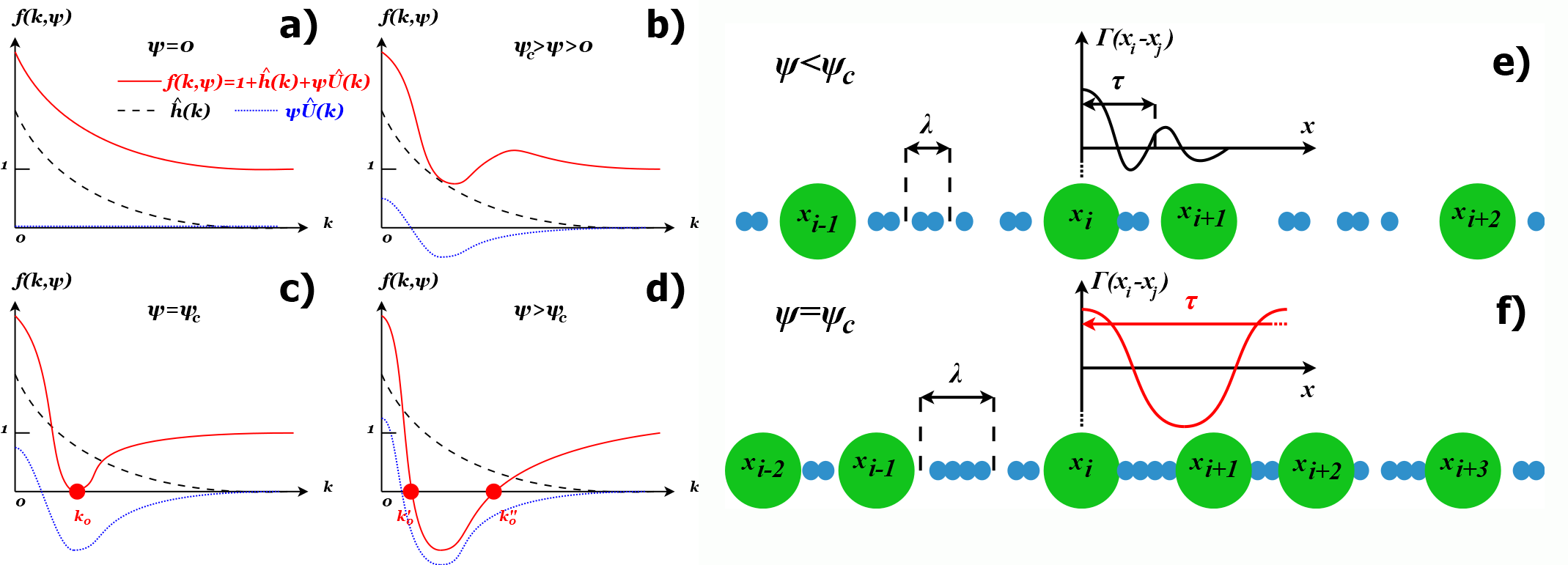}
\caption{a)-d): Root formation in $f(k,\psi)$ (red solid line), i.e. the denominator of Fourier spectrum of friction function $\Gamma(r)$ (see formulas \eqref{eq:Gamma_fin} and \eqref{eq:I}), which is responsible for the SCN-induced arrest. Since the correlation function $\hat h(k)$ (dashed black line) is always a non-negative function and $\hat U(k)$ (dotted blue line) usually has a negative region, $f(k,\psi)$ must be non-negative for small values of $\psi$ (panels a) nad b)). However, there exist a critical value $\psi_c$, such that $f(k_0,\psi_c)=0$  in exactly one point $k_0$ (panel c)). This is the source of the non-integrable singularity in \eqref{eq:Gamma_fin}, leading to the glass-like transition. For $\psi>\psi_c$ (panel d)) the single root $k_0$ splits into two other roots, each now related to an integrable singularity. e)-f): The qualitative origin of the huge collective dissipation length $\tau$ in $\Gamma(r)$. The noise correlation length $\lambda$ is the effective measure for the size of clusters formed by the the smaller particles. e) For a loosely packed system these clusters mediate the influence of the $i$-th bigger particle only to its immediate surroundings. f) For a densely packed system, the influence is mediated over the large distances via the cascade of interactions between the big particles and clusters. \label{fig:cart}}
\end{figure}

The behavior of $f(k,\psi)$ with growing $\psi$ is illustrated in the Fig. \ref{fig:cart}a-d. We demand that $ \hat h(k)\ge 0$ (which is necessarily true as $\textbf{H}$ must be the positive definite matrix) and $\hat h(k)\le1$, i.e. the Fourier spectrum of $H(r)$ is limited by the spectrum of the delta-like correlations. Under these assumptions \eqref{eq:cond} has a solution, provided that $\hat U(k)<0$ for at least some $k\in [0,\pi m/d]$. This is the necessary condition. For $\psi\simeq 0$ we have $f(k,\psi)>0$ regardless of $\hat U(k)$ and $I(k,\psi)$ is non-negative (Fig.\ref{fig:cart}a and \ref{fig:cart}b). As $\psi$ increases, one must encounter the critical value $\psi_c$, such that:
 \begin{align}
f(k_0,\psi_c)=0&&\partial_{k_0}f(k_0,\psi_c)=0 \label{eq:cond2}
 \end{align}
 so $k_0$ is the global minimum of $f(k,\psi_c)$ for $k\in[0,\pi m/d]$, i.e. we have $f(k,\psi_c)>0$ everywhere except for $k_0$ (Fig. \ref{fig:cart}c). Therefore, as $\psi\to \psi_c$, $I(k,\psi)$ develops a maximum at $k_0$, which becomes infinite and non-integrable for $\psi=\psi_c$. Indeed, expanding $I(k,\psi)$ about $k_0$, one can show that $\pmb{\Gamma}_{ii}\propto (\psi-\psi_c)^{-1/2}$ (see Appendix \ref{app:exponent}) in the direct vicinity of $\psi_c$. At this point $\pmb{\Gamma}_{ii}$ dominates in the dissipation, thus this divergence heralds the molecular arrest. As the disorder is also inherently assumed in this theory, we associate this singularity with glass-like transition. The coordinates of the critical point, i.e. $k_0$ and $\psi_c$ can be determined from \eqref{eq:cond2}. For $\psi>\psi_c$ (Fig.\ref{fig:cart}d) the solution of $f(k,\psi)=0$ bifurcates, i.e. there are two solutions $k_0'$ and $k_0''$ such that $k_0'<k_0<k_0''$. The singularities at these points are of the removable type, so \eqref{eq:Gamma_fin} once again can be integrated to a finite value. However, for $k\in[k_0',k_0'']$ we encounter $f(k,\psi)<0$, which causes the Fourier spectrum of $\pmb{\Gamma}$ to be partially negative. This means that $\pmb{\Gamma}$ is no longer a positive-definite matrix i.e. its purely dissipative character is violated. Since we also know that once the system gets arrested for $\psi_c$, it should stay as such for $\psi>\psi_c$, this means that \eqref{eq:Gamma_fin} becomes unphysical in this regime. However, there exists another solution to \eqref{eq:FPE0} which was not discussed yet, i.e. for $\Lambda_{K,k}=1$ (constant spectrum) we also obtain $\pmb{\Gamma}_{ii}\to +\infty$. Therefore, we claim that for $\psi>\psi_c$ the system spontaneously switches from the solution \eqref{eq:Gamma_fin} to infinite viscosity. This discontinuity supports the arrested state for the higher packings. 

Finally, we can also discuss the behaviour of the non-diagonal terms in the matrix $\pmb{\Gamma}$, which vary with distance. For all $i\neq j$  they have the same functional form, so we will denote them in general as $\pmb{\Gamma}_{ij}(r)=\Gamma(r)$, also in contrast to $\pmb{\Gamma}_{ii}=\Gamma(0)$. As $\psi\to \psi_c$, the integral in \eqref{eq:Gamma_fin} is dominated by the contribution from the vicinity of $k_0$. Thus, one might expand the integrand around this value to obtain the analytical approximations for $\Gamma(r)$ (see Appendix \ref{app:exponent} for details). This leads to the following results:
\begin{align}
\pmb{\Gamma}_{ij}(r)\simeq \pmb{\Gamma}_{ii} e^{-\frac{|r|}{\tau}}\cos k_0r&&\pmb{\Gamma}_{ii}\simeq \frac{\gamma \hat h(k_0)(2+\psi \hat U(k_0)) }{\sqrt{2(\psi_c-\psi)f''(k_0,\psi)|\hat U(k_0)|}} &&\tau=\sqrt{\frac{f''(k_0,\psi)}{2(\psi_c-\psi)|\hat U(k_0)|}}\label{eq:non_diag_gamma}
\end{align}
First, one might notice that, globally, the scale of dissipation is governed by the diagonal terms $\pmb{\Gamma}_{ii}$. Further, for $\psi<\psi_c$ (as soon as $f''(k_0,\psi)>0$) the non-diagonal terms decay exponentially with range. This decay is controlled by $\tau$, i.e. the characteristic length of collective dissipation, whose nature is of crucial meaning to the SCN-driven dynamics. While for $\psi<\psi_c$ this length is finite (which ensures the locality of dissipation), it diverges at $\psi=\psi_c$. In this situation, the strongly oscillatory behaviour dominates ($\pmb{\Gamma}_{ij}(r)\simeq\pmb{\Gamma}_{ii}\cos k_0 r$) and the dissipation becomes scale-free ($\tau\to+\infty$). This means that the dissipation is highly non-local, i.e. every two, arbitrarily distant, particles affect each others' dissipation in a significant manner. Remarkably, the transition into this highly cooperative state manifests in the dynamics, not structure, as the noise correlation length $\lambda$ remains finite both below and above $\psi_c$. Thus, the SCN-induced arrest possesses a divergent length-scale that does not manifest in the system ordering. 

The intuitive understanding of the divergence in $\tau$ is presented in the Fig.\ref{fig:cart}e-f. Let us assume that the thermal noise is correlated over a certain length $\lambda$. In the microscopic sense, this means that the smaller particles move together in the clusters of size $\lambda$. To certain extent, one can imagine these clusters as the solid, finite size objects, mediating the interactions between the larger particles. When a large particle collides with these clusters, the disturbance transfers to its immediate surroundings. However, at $\psi<\psi_c$, the chance that another large particle is affected is low and so the disturbance dies out. Thus, $\tau$ has a finite range (see Fig.\ref{fig:cart}e). However, near $\psi_c$ and above, the large particles are abundant enough so that the disturbance can easily transfer over the great distances via the sequence: particle$\to$cluster$\to$particle$\to$cluster etc. This translates into the divergent behaviour of $\tau$ (Fig.\ref{fig:cart}f). As we will show in the next section, $\psi$ and $\lambda$ work interchangeably, i.e. higher $\lambda$ requires lower $\psi_c$. This is understandable, as the larger clusters can affect particles over greater distances.

Finally, we can also address the mean square displacement (MSD) of the larger particles in the vicinity of the transition. This problem is analytically tractable if we assume that the interactions are relatively short-ranged and the particles are sparse enough, so effectively $F(x_i-x_j)\simeq0$ for all $i$ and $j$. In this case, the equation of motion for $x_i$ reduce to:
\begin{align}
\dot x_i=\sum_{k=-(M-1)/2}^{(M-1)/2} \sigma Q_{ik} \Lambda_{\Gamma,k}^{-1}\sqrt{\Lambda_{H,k}}\eta_k(t)=\sigma \sqrt{\mathcal{D}_i} \omega(t)&&\mathcal{D}_i=\sum_{k=-(M-1)/2}^{(M-1)/2} Q_{ik}^2\Lambda_{\Gamma,k}^{-2}\Lambda_{H,k} \label{eq:MSD1}
\end{align}
where we made use of the fact that the sum of Gaussian variables $\eta_k(t)$ can be replaced by a single Gaussian variable $\omega(t)$ with a properly rescaled variance. In the thermodynamic limit, this variance reads:
\begin{equation}
\sigma^2 \mathcal{D}_i =\frac{2}{\pi \beta  \gamma \rho a_H} \int_{0}^{\frac{\pi m}{d}}dk \frac{(1+\hat h(k)+\psi \hat U(k))^2}{\hat h(k)(2+\psi \hat U(k))^2}
\end{equation}
which is independent of $x_i$. Thus, we can instantly solve \eqref{eq:MSD1} and obtain MSD:
\begin{align}
x_i(t)=x_i(0)+\sigma \sqrt{\mathcal{D}_i}\int_0^tds \omega(s)&&<(x_i(t)-x_i(0))^2>= \sigma^2\mathcal{D}_i t
\end{align}
Such MSD shows that in our model the diffusion of larger particles is normal, up to the transition point $\psi_c$. This is consistent with the known characteristics of the supercooled state, below the glass transition, which was confirmed for e.g. polydisperse hard-sphere colloids\cite{bib:mass_sims} and micellar solutions \cite{bib:likos_diff}. Thus, the SCN-induced arrest apparently corresponds to the transition between the supercooled and the glassy state. The diffusion coefficient $\sigma^2 \mathcal{D}_i$ decreases with growing $\psi$, due to the influence of $f(k,\psi)$ on its integrand.

\section*{Examples of application: soft particles and hard spheres} \label{sec:example}
We can now conveniently illustrate the mechanism of SCN-induced arrest in the two types of systems, soft particles and hard spheres, to identify its universal characteristics. This allow us to eventually discuss the whole range of the binary mixture dynamics grasped by our model, in the following section. The first, and possibly the simplest case, are the particles interacting via the Dirac-delta potential and affected by the exponentially correlated noise, i.e.:
\begin{align}
H(r)=\frac{e^{-\frac{|r|}{\lambda d}}}{\lambda }\label{eq:exp_corr}&&U(r)=\epsilon \delta\left( \frac{|r|-d}{d}\right)
\end{align}
where $d$ is the particle diameter, $\lambda$ is the (dimensionless) ratio between the particle size and the correlation length and $\epsilon$ is the interaction energy. We choose the exponential correlation function as it is observed in supercooled systems\cite{bib:doliwa,bib:weeks}, where $\lambda$ is known to grow with density. Thus, this correlation function corresponds to a highly packed thermal bath, on the verge of vitrification. It also allows us to analyse how the system responds to a well-defined noise correlation length. This would not be possible for the power-law correlations, which originate from the hydrodynamic interactions\cite{bib:Deutch}. The potential $U(r)$ represents the extremely soft particles, which interact only when their 'surfaces' touch and can also penetrate each other almost without any energetic penalty. While this potential is not a realistic one, it provides the clearest demonstration (also in the mathematical sense) of the transition induced by SCN. In other words, the dynamics diminish not due to the direct interactions of the particles (as they can rearrange almost freely), but because of their collective interactions with the thermal bath. In our model $H(r)$ satisfies $\lim_{\lambda\to0}H(r)=\delta(r/d)$, in accordance with the initial assumption $\lim_{\lambda\to0}H(r)=\frac{a_H}{\rho}\delta(r)$. This allows us to determine $a_H$. Eventually, the appropriate Fourier transforms read:
\begin{align}
\hat U(k)=2d\epsilon\cos kd&&\hat h(k)=\frac{1}{1+(\lambda k d)^2}&&a_H=\rho d
\end{align}
Inserting these formulas into \eqref{eq:Gamma_fin} and substituting $z=kd$, we obtain:
\begin{equation}
\frac{\pmb{\Gamma}_{ii}}{\gamma}=\frac{1}{\pi}\int_0^{\pi m}dz\frac{\frac{1}{1+z^2\lambda^2}(2+2\tilde \psi \cos z)}{1+\frac{1}{1+z^2\lambda^2}+2 \tilde \psi \cos z} \label{eq:example}
\end{equation}
where $\tilde \psi=\beta \epsilon \rho d$ is the rescaled temperature-packing parameter. We choose $\epsilon=\beta^{-1}$ (interaction energy matches the thermal fluctuation scale), so $\tilde \psi=\rho d$ becomes the actual volume fraction occupied by particles. The system is fully described by $\tilde \psi$, $\lambda$ and the cut-off parameter $m$. 

\begin{figure}[ht]
\includegraphics[width=\columnwidth]{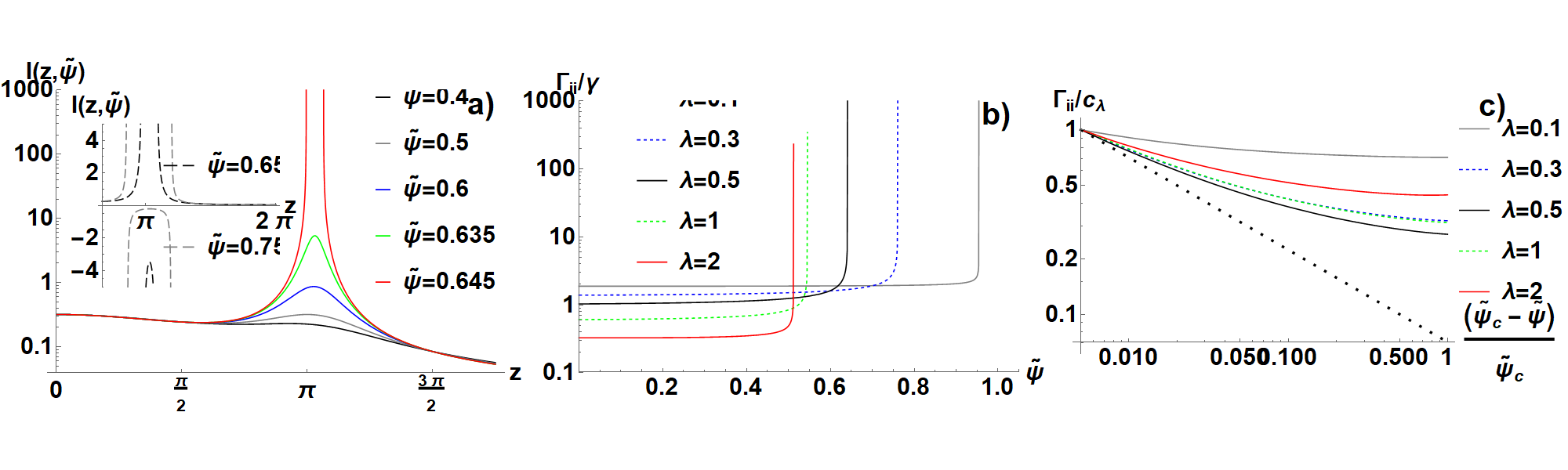}
\caption{a) The Fourier spectrum of friction coefficient $\Gamma(r)$ in the soft interaction case (i.e. the function $I(z,\tilde \psi)$, see equations \eqref{eq:I} and \eqref{eq:example}), as the function of wave number $z$ and the temperature-packing parameter $\tilde\psi$, for $\lambda=0.5$ and $m=2$. The symmetric, non-integrable singularity emerges as $\tilde \psi$ approaches the critical value $\tilde \psi_c\simeq0.645$. Inset: $I(z,\tilde \psi)$ continued to $\tilde \psi>\tilde \psi_c$ shows a negative spectrum. b) The dominant, diagonal friction coefficient $\pmb{\Gamma}_{ii}/\gamma$ as a function of $\tilde \psi$, showing a divergence at the critical point $\tilde \psi=\tilde \psi_c$ (see Fig \ref{fig:crit_psi}a). This divergence marks the transition into the arrested state and it is a consequence of the singularity illustrated in the panel a). Different curves illustrate the dependence on the correlation parameter $\lambda$, $m=2$. c) The log-log plot of $\pmb{\Gamma}_{ii}/c_{\lambda}$ as the function of $(\tilde\psi_c-\tilde\psi)/\tilde\psi_c$, illustrating a power-law divergence. $c_{\lambda}$ is chosen to normalize the data, i.e. $c_\lambda=\pmb{\Gamma}_{ii}$ for $(\tilde \psi-\tilde \psi_c)/\tilde \psi_c=0.005$. The dotted line corresponds to the theoretical power-law approximation $\pmb{\Gamma}_{ii}\propto(\tilde\psi_c-\tilde\psi)^{-1/2}$. \label{fig:gamma}\label{fig:integrand}}
\end{figure}
\begin{figure}[ht]
\includegraphics[width=\columnwidth]{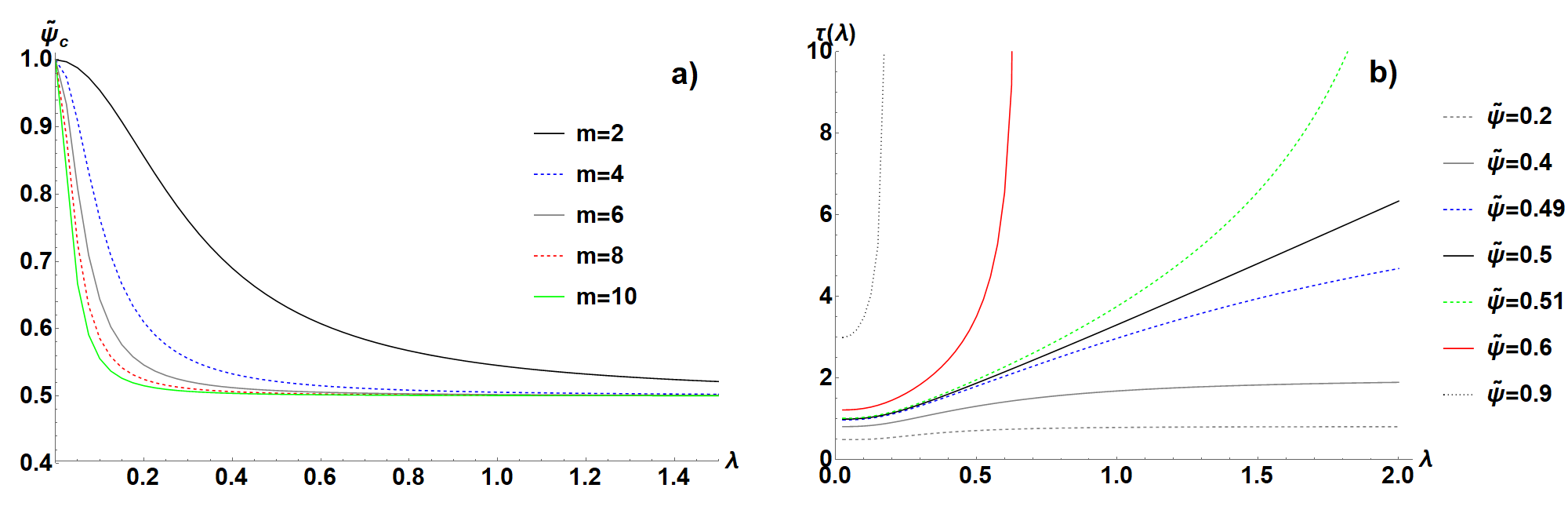}
\caption{a) The critical temperature-packing parameter $\tilde \psi_c$ as a function of correlation parameter $\lambda$, for several values of the cut-off parameter $m$ in the soft-interactions case. $\tilde \psi_c$ marks the glass-like transition and it is determined from \eqref{eq:cond2}. b) $\tau$, the characteristic length of collective dissipation (given by \eqref{eq:non_diag_gamma}), as the function of $\lambda$, for $m=2$ and several values of $\tilde \psi$. $\tilde \psi=0.5$ separates two regions: for $\tilde \psi<0$ $\tau$ grows sub-linearly with $\lambda$, while for $\tilde \psi>0.5$ $\tau$ develops a singularity at a finite $\lambda$, satisfying the equation $\tilde \psi=\tilde \psi_c(\lambda)$ (see panel a) ). \label{fig:crit_psi}\label{fig:tau}}
\end{figure}
\begin{figure}[ht]
\includegraphics[width=\columnwidth]{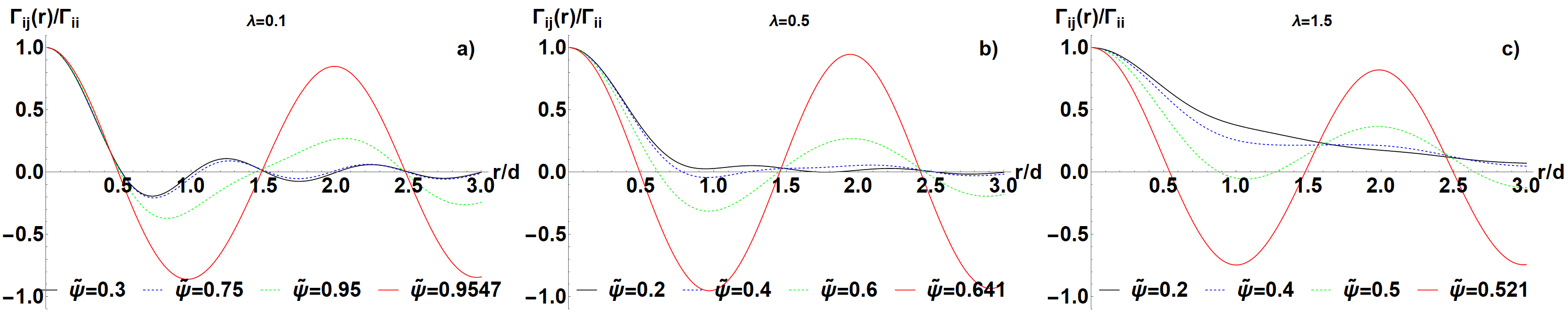}
\caption{The spatial dependence of the non-diagonal friction coefficient $\pmb{\Gamma}_{ij}(r)$ normalized to $\pmb{\Gamma}_{ii}$ for soft interactions. These coefficients inform how many neighbouring particles affect each others dissipation. The results are generated for $m=2$ and $\lambda=0.1$ (a), $\lambda=0.5$ (b) and $\lambda=1.5$ (c). The functions have a decaying character, with the effective range growing with $\lambda$, as predicted by \eqref{eq:non_diag_gamma}. The oscillatory behaviour develops with the increasing $\tilde \psi$, dominating in the vicinity of the critical value for the glass-like transition. \label{fig:gamma_ij}}
\end{figure}


Fig. \ref{fig:integrand}a illustrates how the integrand of \eqref{eq:example} evolves with the growing $\tilde \psi$, for $\lambda=0.5$. One can clearly notice the build-up of the infinite maximum for $\tilde\psi<\tilde\psi_c\simeq 0.645$ and, later, the splitting of this maximum for $\tilde \psi>\tilde \psi_c$. In Fig. \ref{fig:crit_psi}a we present the influence of $\lambda$ and $m$ on $\tilde \psi_c$. From \eqref{eq:Gamma_fin}, it follows that for the non-correlated system ($\lambda=0$, $\hat h(k)=1$) there is no transition and the system obeys ordinary diffusion. However, Fig. \ref{fig:crit_psi}a shows that for $\lambda\to 0^+$ the arrest persists until $\tilde \psi_c=1$ (the maximal packing). This shows that there is a discontinuity at $\lambda=0$, i.e. even the infinitesimally small spatial correlations make the transition possible. Inspection of \eqref{eq:Gamma_fin} shows that this conclusion holds for any $H(r)$ that tends to the delta behaviour with $\lambda\to0$. On the other hand, as $\lambda$ increases, $\tilde \psi_c$ decreases and quickly saturates at 0.5. This means that below critical packing even the extremely long correlations cannot arrest the system via the FDR-embedded mechanism. However, this result indicates a change in the slow-down mechanism and we will scrutinize this issue in the next section. Once again, \eqref{eq:Gamma_fin} shows that such limiting critical packing should exist for all $H(r)$ satisfying $\lim_{\lambda\to+\infty}H(r)=const$ (thus $\lim_{\lambda\to +\infty}\hat h(k)=0$ for $k\neq 0$), though its exact value should depend on $\hat U(k_0)$. The final observation is that for the cut-off $m\to+\infty$ we obtain $\tilde \psi_c\to 0.5$ for all $\lambda>0$. This behaviour is specific to our soft potential. In the Fig. \ref{fig:gamma}b the behaviour of $\pmb{\Gamma}_{ii}$ is presented, which clearly shows the divergence of viscosity in the vicinity of the critical point. The jump is at least three orders of magnitude. In the Fig. \ref{fig:gamma}c the same data are also shown in the collapsed form on the log-log plot, as the function of $(\tilde\psi_c-\tilde\psi)/\tilde\psi_c$. This allows us to compare the actual behavior of $\pmb{\Gamma}_{ii}$ with the approximated theoretical power-law $(\tilde\psi_c-\tilde\psi)^{-1/2}$. The plot shows that the numerical data decay slower than this power-law and the convergence is not achieved until very close to the singularity.

 Finally, in the Fig. \ref{fig:gamma_ij}a-c the non-diagonal dissipative terms $\pmb{\Gamma}_{ij}(r)/\pmb{\Gamma}_{ii}$ are plotted in the real space for $m=2$ and $\lambda=0.1$, 0.5 and $1.5$. Indeed, these functions evolve with the growing $\tilde \psi$ as predicted by the formula \eqref{eq:non_diag_gamma}. For the low packing fraction $\pmb{\Gamma}_{ij}(r)/\pmb{\Gamma}_{ii}$ has a decaying character and its effective range grows with $\lambda$. As $\tilde\psi$ increases, the oscillations become stronger and for $\tilde \psi\simeq\tilde \psi_c$ they completely dominate. The period of these critical oscillations proves practically independent from $\lambda$, but this is not universal and might be related to the type of $H(r)$ we use. In the Fig. \ref{fig:tau}b the characteristic length-scale $\tau$ of the collective dissipation is presented as a function of $\lambda$. This figure shows how the thermal noise correlations translate into correlations in dynamics. The behaviour is highly non-trivial because $\tilde \psi_c$ in formula \eqref{eq:non_diag_gamma} is also the function of $\lambda$. Two regimes of $\tau$ can be distinguished: for $\tilde \psi<0.5$ it grows sub-linearly with $\lambda$ reaching $\tau\to+\infty$ as $\lambda\to +\infty$. The $\tilde\psi=0.5$ is the limiting case in which $\tau$ becomes a linear function of the large $\lambda$. However, for $\tilde \psi>0.5$ the singularity in $\tau$ emerges, i.e. $\tau\to+\infty$ as $\lambda$ approaches the solution of $\tilde\psi=\tilde \psi_c(\lambda)$, where $\tilde\psi_c(\lambda)$ is shown in Fig. \ref{fig:crit_psi}a. This behaviour is a manifestation of the SCN-induced transition.

We discuss the hard-sphere (HS) interactions, now. This potential can be understood as an infinite energy barrier when two particles are closer than $d$ and 0 otherwise. Such interaction has no well defined Fourier transform. However, one can interpret the HS potential in a slightly different manner, i.e. as an interaction that makes the stationary distribution $P_s$ (as given by \eqref{eq:Ps}) equal to 0 if any pair of particles satisfies $|x_i-x_j|<d$. Let us define the following function:
\begin{equation}
U_{HS}(r)=-\frac{\epsilon}{\beta}\ln[\theta(|r|-d)]
\end{equation}
where $\theta(x)$ is the Heavyside step function and $\epsilon>0$ is a dimensionless scaling constant that we will tune later. Inserting this $U_{HS}(r)$ into $P_s$ we obtain:
\begin{gather}
P_s\propto\prod^N_{\substack{n,m\\ n> m}}\theta^{\epsilon}(|x_n-x_m|-d)e^{-\frac{\beta}{2}\sum_{i,j}U(x_i-x_j)}
\end{gather}
where $U(r)$ are the soft potentials (if any). This shows that $U_{HS}(r)$ acts as the HS potential for any $\epsilon>0$. The main challenge now is to determine the Fourier transform of $U_{HS}(r)$. In the following calculations we understand that $\theta(|r|-d)=1-\theta(r+d)+\theta(r-d)$. The first step is to apply the integration by parts:
\begin{equation}
\begin{split}
\hat U_{HS}(k)&=-\frac{\epsilon}{\beta}\int_{-\infty}^{+\infty}dr e^{-ikr}\ln[\theta(|r|-d)]=\frac{\epsilon}{\beta}\int_{-\infty}^{+\infty}dr\frac{e^{-ikr}}{ik}\frac{\delta(r+d)-\delta(r-d)}{\theta(|r|-d)}
\end{split}
\end{equation}
Let us now suppose that there exists a function satisfying:
\begin{equation}
\delta(r+d)-\delta(r-d)=\frac{\theta(|r|-d)}{2\pi}\int_{-\infty}^{+\infty}dk' e^{ik'r} \hat f(k') \label{eq:hs1}
\end{equation}
then $\hat U_{HS}(k)=\frac{\epsilon}{ik\beta}\hat f(k)$. Indeed, this is true and in Appendix \ref{app:HS} we show that the solution reads:
\begin{equation}
\hat U_{HS}(k)=\frac{4\epsilon}{\beta}\frac{\sin kd}{k}
\end{equation}
which is, in fact, equivalent to the Fourier transform of the rectangular signal. The final step is to tune $\epsilon$. We know that in one dimension, in the presence of the infinitesimally small correlations, $\lambda \to 0^+$ (so $\hat h(k)\to1$), the system of hard spheres must jam for the packing fraction $\psi=\rho d = 1$. According to \eqref{eq:cond}, this transition is possible, provided that $\epsilon$ satisfies:
\begin{equation}
2+\frac{4\epsilon \sin k_0d}{k_0d}=0
\end{equation}
where $k_0d$ corresponds to the global minimum of the sinc function, i.e. $\frac{\sin k_0d}{k_0d}\simeq-0.217$. Therefore $\epsilon\simeq 2.3$.
 
\begin{figure}[ht]
\includegraphics[width=\columnwidth]{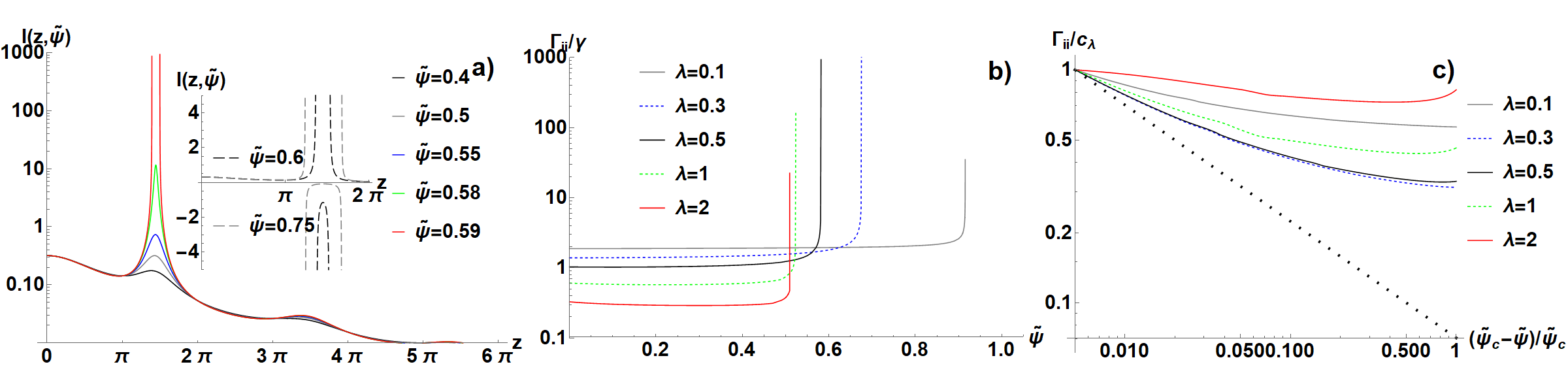}
\caption{a) The Fourier spectrum of $\Gamma(r)$ in the hard-sphere case, i.e. the function $I(z,\tilde \psi)$ (see formula \eqref{eq:hs2}), for $\lambda=0.5$ and $m=2$. The singularity emerges at the critical packing $\tilde\psi\simeq 0.59$. b) The dominant friction coefficient $\pmb{\Gamma}_{ii}$ for hard spheres as a function of the packing fraction $\tilde \psi$ for several values of $\lambda$ and $m=2$. c) The same coefficient collapsed on the log-log plot and compared to the theoretical prediction $(\tilde \psi_c-\tilde \psi)^{-1/2}$ (dotted line). $c_\lambda$ is chosen to normalize the data, i.e. $c_\lambda=\pmb{\Gamma}_{ii}$ for $(\tilde \psi-\tilde \psi_c)/\tilde \psi_c=0.005$.\label{fig:hs1}}
\end{figure}

\begin{figure}[ht]
\includegraphics[width=\columnwidth]{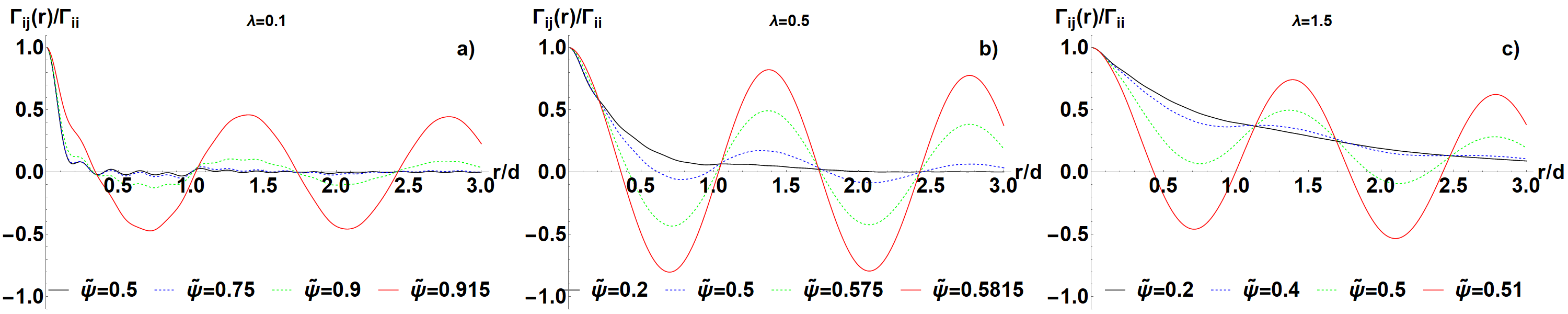}
\caption{The non-diagonal friction coefficient $\pmb{\Gamma}_{ij}(r)/\pmb{\Gamma}_{ii}$ in the hard-sphere case for $m=10$ and $\lambda=0.1$(a), $\lambda=0.5$(b) and $\lambda=1.5$(c). In all cases the oscillatory character emerges as $\tilde\psi\to \tilde \psi_c$. For short correlations ($\lambda=0.1$) the additional oscillatory structure manifests in the lower packing regime. \label{fig:hs4}}
\end{figure}

With these considerations at our disposal, we conclude that $\pmb{\Gamma}_{ii}$ for the purely hard-sphere system with exponential correlations given by \eqref{eq:exp_corr} reads:
\begin{equation}
\frac{\pmb{\Gamma}_{ii}}{\gamma}=\frac{1}{\pi}\int_0^{\pi m} dz \frac{\frac{1}{1+\lambda^2z^2}(2+4\tilde\psi\epsilon\frac{\sin z}{z})}{1+\frac{1}{1+z^2\lambda^2}+4\tilde\psi\epsilon\frac{\sin z}{z}}\label{eq:hs2}
\end{equation}
where $\tilde \psi =\rho d$ (packing fraction). Fig. \ref{fig:hs1}a presents the behaviour of the integrand in \eqref{eq:hs2} with a growing packing fraction for $\lambda=0.5$. As previously, one can observe the build-up of the singularity as the packing approaches $\tilde \psi_c\simeq 0.59$. Fig. \ref{fig:hs2}a shows the dependence of the critical packing $\tilde \psi_c$ on the correlation length. The HS system proves almost completely insensitive to the cut-off factor $m$. As soon as $m\ge2$, all the curves collapse on the same plot, which is in stark contrast with the soft-potential case. This is because $\hat U_{HS}(k)$ is a fast decaying function, with its global minimum relatively close to $k=0$. However, in the long correlation case, $\tilde \psi_c$ saturates at 0.5, similarly to the soft interaction case. In Fig. \ref{fig:hs1}b and \ref{fig:hs1}c, $\pmb{\Gamma}_{ii}$ is plotted as the function of $\tilde \psi$. Again the convergence to the power-law behaviour $(\tilde \psi_c-\tilde \psi)^{-1/2}$ is achieved only as $\tilde\psi\to\tilde\psi_c$. Finally, in the Fig. \ref{fig:hs4}a-c the non-diagonal dissipative terms $\pmb{\Gamma}_{ij}(r)/\pmb{\Gamma}_{ii}$ are plotted. Their general character resembles the soft-interaction case and agrees with the predictions of \eqref{eq:non_diag_gamma}. However, the hard-sphere case reveals an additional oscillatory structure for the low packing and short correlations ($\lambda=0.1$). It is also in this regime that the dominant oscillations emerge only very close to the transition. For the longer-ranged correlations the evolution is more gradual. Fig. \ref{fig:hs2}b presents the behaviour of $\tau$, which proves completely analogous to the soft-interaction case.

The general theoretical predictions regarding the transition into the SCN-induced arrest, in particular the behaviour of $\pmb{\Gamma}_{ii}$ and $\pmb{\Gamma}_{ij}(r)$, are fully confirmed in both systems analysed. The poorest agreement is obtained for the power law exponent, which seems to be adequate only very close to $\tilde\psi_c$. Although the systems of ultra soft particles and hard-spheres are very different, many similarities are shared. Our calculations pinpoint a few universal features of the transition such as: the minimal packing required for the transition, the persistence of the transition for $\lambda\to 0^+$ and the oscillatory behaviour of $\pmb{\Gamma}_{ij}(r)$ as $\tilde \psi \to \tilde \psi_c$. The differences between soft and hard particles are mainly quantitative, though the direct comparison is hindered as the behaviour of the soft particles varies considerably with the cut-off $m$. However, at least for the higher values of $m$, the system of soft particles seems more prone to arrest. It manifests in the generally lower values of $\tilde \psi_c$ and $\lambda$ necessary for the transition. This is also reflected in the shape of $\pmb{\Gamma}_{ij}(r)$. While the type of the potential has a minor effect on this function close to the transition, it strongly affects the low-$\tilde \psi$ and low-$\lambda$ regime. In this case, $\pmb{\Gamma}_{ij}(r)$ for hard spheres is visibly concentrated around $r=0$, so the friction in this regime is effectively Stokesian, while for soft particles $\pmb{\Gamma}_{ij}(r)$ still has a considerable range. This might be caused by the fact that two soft particles can penetrate each other without major energetic penalty, thus they can feel the length-scale of noise correlations $\lambda$, even if it is smaller than the particle size. This is not possible for hard spheres, so the collective effects in this system are less pronounced.

\section*{Regimes of SCN-driven dynamics vs. different  modes of arrest in binary mixtures}
We will now review the above results from the perspective of binary mixture dynamics, with emphasis on the behaviour of the larger (observed) particles. Let us remind that $\lambda$ reflects the phase-state of the thermal bath in our model. We expect that $\lambda$ grows monotonically with increasing packing fraction of the thermal bath. More specifically, when $\lambda$ is comparable with the diameter of the smaller particles, it corresponds to the liquid thermal bath (only nearest neighbours are correlated), while $\lambda$ ranging over a few diameters of smaller particles means that the thermal bath is in the super-cooled or arrested state already. In general, in binary mixtures, the state of the thermal bath is not independent from the state of the bigger species, as it is the total packing fraction of the system (small + large particles) that matters. However, in our case, it is convenient to discuss various regimes, treating $\lambda$ as an independent variable. 

First of all, we may consider very low concentration of the smaller particles, which corresponds to the $\lambda\to 0^+$ limit. In this case, the FDR for SCN predicts arrest only for an extremely high packing fraction of the larger particles, $\tilde \psi\simeq 1$. However, in fact, $\lambda\to 0^+$ corresponds to a mono-dispersed colloid with ordinary Brownian dynamics. Such systems can undergo structural arrest directly due to the inter-particle forces $F(x_i-x_j)$. This mechanism is likely to overshadow the SCN-induced arrest, especially for the long-range forces, as it is likely to require a critical $\tilde \psi_{SA}$ lower than 1. Thus, the structural arrest is embedded in our model in the interaction terms, alongside the SCN-induced arrest.

\begin{figure}[ht]
\includegraphics[width=\columnwidth]{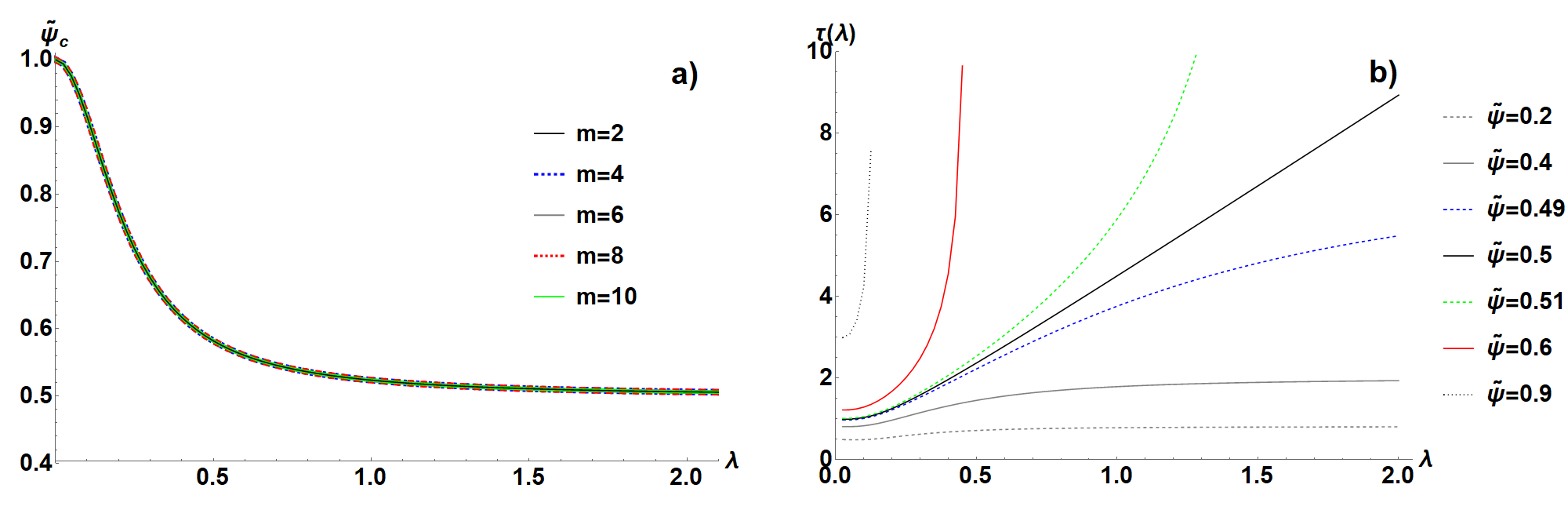}
\caption{Hard sphere case: a) The dependence of critical packing fraction $\tilde \psi_c$ on the correlation length $\lambda$. $\tilde \psi_c$ proves virtually insensitive to the change in the cut-off factor $m$, unlike in the soft interaction case (compare Fig. \ref{fig:crit_psi}). b) $\tau$, the characteristic length of the collective dissipation (given by \eqref{eq:non_diag_gamma}), as the function of $\lambda$, for several values of packing $\tilde \psi$. The results are analogous to the soft interaction case (see Fig. \ref{fig:tau} for details) \label{fig:hs2}}
\end{figure}

For intermediate concentrations of the smaller particles we expect a finite value of $\lambda$. For $\tilde \psi<\psi_c$ the system remains liquid. For $\tilde \psi\ge \tilde \psi_c$, the SCN-induced arrest is expected to emerge, which might be understood as a cooperative phase, sustained by the interactions between the both types of particles. The curves on the $\lambda$-$\tilde \psi_c$ plot (Fig. \ref{fig:crit_psi}a and \ref{fig:hs2}a) show the separation line between these two phases. The SCN-induced arrest might also compete with the structural arrest in this regime, especially for higher packing, though this depends on the structural details of a particular system.

Finally, for the high packing fraction of the smaller particles, $\lambda$ becomes very large. This leads to another type of dynamics, which can be conveniently analysed as the $\lambda\to+\infty$ limit. In this case, the correlation matrix reads $\textbf{H}_{ij}=\sigma^2$, i.e. all its elements are equal. Thus, we have virtually $\xi_i(t)=\xi_j(t)$ for every pair of $i$ and $j$. Also, it means that $\hat h(k)\propto \delta(k)$, so, following \eqref{eq:Gamma_fin}, $\pmb{\Gamma}_{ij}=const.=\gamma_\infty$. If we switch from the absolute positions $x_i$ to the variable $\Delta x_i$, relative to the mass centre of the system, the equations of motion reduce to ${\gamma_{\infty}\sum_j^N \Delta \dot x_j=\sum_j F(\Delta x_i-\Delta x_j)}$. These dynamics have no stochastic driving, so particles relax their initial kinetic energy exponentially fast, before becoming asymptotically immobile. Thus, this state is also arrested, but the mechanism of this effect is different from the SCN-induced arrest: it involves finite dissipation. The transition between the  mobile phase and this type of arrest is gradual, as the relative stochastic driving diminishes with growing $\lambda$.
 
In summary, there are up to four modes of dynamics for bigger particles, which occupy different regions of on the $\lambda$-$\psi_c$ plot: the structural arrest ($\lambda\simeq 0$, $0.5<\tilde \psi<\tilde \psi_c$ ), the SCN-induced arrest (finite $\lambda$, $\psi\ge\psi_c$), the mobile phase (finite $\lambda$, $\tilde \psi<\tilde \psi_c$) and the infinite-$\lambda$ arrest ($\lambda\to+\infty$, $\tilde \psi<\tilde \psi_c$). The experimental phase diagram for binary mixtures also contains four main disordered phases, as mentioned in the introduction: single glass, double glass, asymmetric glass and double fluid\cite{bib:binmix2}. These phases might correspond to the modes of dynamics identified in our model. In this interpretation, the mobile phase is the double fluid and the infinite-$\lambda$ arrest is the asymmetric glass. Further, the double glass phase corresponds to the SCN-induced arrest as it is a cooperative phase. The single glass might correspond either to the SCN-induced arrest too, or to the structural arrest. This is likely to depend on the parameters of a particular system. The four phases embedded in the SCN-driven dynamics are also consistent with the re-entrant transition, i.e. the melting of the binary mixture as the bigger particles are replaced by the smaller ones, with the total packing fraction for both types kept constant\cite{bib:binmix2}. Looking at Fig. \ref{fig:crit_psi}a and \ref{fig:hs2}a, one can notice that decreasing $\tilde \psi$ moves the system downwards on the $\lambda$-$\tilde \psi_c$ plot. Thus, melting should occur as this transformation crosses the $\tilde \psi_c(\lambda)$ line. The compensating increase in the number of smaller particles corresponds to the growing $\lambda$ and moves the system to the right on the $\lambda$-$\tilde\psi_c$ plot. The shape of the $\tilde \psi_c(\lambda)$ curve is such that this transformation can either move the system back to the SCN-induced arrest or towards the infinite-$\lambda$ arrest. In both cases, this would result in the re-entrant transition.

Concluding this section, despite being a relatively simple model, the SCN-driven Langevin dynamics qualitatively cover a few major phenomena encountered in binary mixtures. However, this phase behaviour might also include e.g. gel phases\cite{bib:binmix2}, crystallization\cite{bib:binmix_cryst,bib:binmix3} and demixing effects\cite{bib:majka}, which are not addressed here. A more in-depth discussion would require a three dimensional model, employing a more specific noise correlation function, i.e. one that explicitly provides the physical dependence between $\lambda$ and the microscopic parameters of the thermal bath. This will be pursued elsewhere, as this article is focused on the most general properties of the SCN-based model.



\section*{Relation to the mode coupling theory}\label{sec:MCT}
The SCN-induced arrest resembles the ideal glass transition described by MCT. This is not surprising, as both theories stem from a common concept. In the usual derivation of MCT, there appears a certain intermediate equation (see equation (25) in  Reichman and Charbonneau (2005)\cite{bib:mct}) that explicitly involves the fluctuating force and the friction memory kernel. This is, in fact, the Generalized Langevin Equation. From this moment, abandoning the standard MCT route in favour of the stochastic approach and adopting the assumptions of SCN, one arrives at the starting point of this paper, which is \eqref{eq:Lan}. This means that both theories can be compared in greater detail.

 The main similarity is that both theories predict a sharp transition, with power-law divergence in the viscosity. In MCT, the critical exponent reads 1.765, as estimated by Leutheusser \cite{bib:leutheusser}, while our model estimates 0.5. Another similarity is that both theories predict a divergent correlation length at the critical point. The main result of MCT is the evolution equation for the dynamic density correlation function, defined as:
\begin{equation}
\mathcal{C}_k(t)=\left< N^{-1}\sum_{i,j}\exp\left[i\frac{2\pi k}{L}\left(x_i(0)-x_j(t)\right) \right]\right>\label{eq:def_C}
\end{equation}
MCT is known to predict the two-step relaxation of $\mathcal{C}_k(t)$, where the $\beta$-relaxation corresponds to the short-to-intermediate time and the long-time behaviour is called the $\alpha$-relaxation. Vitrification manifests in the extension of $\alpha$-relaxation to extremely long times, saturating at a constant value, which marks the transition from ergodic to non-ergodic dynamics \cite{bib:mct}. 

Within our theory, we can also predict how $\mathcal{C}_k(t)$ evolves. First, we need to introduce the variable:
\begin{equation}
y_k=\frac{1}{\sqrt{N}}\sum_i^N\left(\cos \frac{2\pi k x_i}{L}-\sin \frac{2\pi k x_i}{L}\right)\label{eq:y_k}
\end{equation}
Under this transformation, the system of Langevin equations \eqref{eq:Lan} translates into (see Appendix \ref{app:mct}):
\begin{equation}
\gamma a_{\Gamma}\Lambda_{\Gamma,k}\dot y_k=-\rho \hat U_k y_k-\frac{L}{2\pi k}\sigma \Lambda_{H,k}^{1/2}\eta_k \label{eq:Lan_y}
\end{equation}
In Appendix \ref{app:mct} we show how $y_k$ and $\mathcal{C}_k(t)$ are related to each other, hence, the above equation also leads to the evolution of $\mathcal{C}_k(t)$:
\begin{equation}
\gamma a_{\Gamma}\dot{\mathcal{C}}_k(t)+\frac{\rho\hat U_k}{\Lambda_{\Gamma,k}} \mathcal{C}_k(t)=0 \label{eq:C}
\end{equation}
This result can be now compared with the standard equation provided by MCT\cite{bib:mct,bib:leutheusser}:
\begin{equation}
\ddot{\mathcal{C}}_k(t)+\alpha_1 \int_0^td\tau \mathcal{K}(k,t-\tau)\dot{\mathcal{C}}_k(\tau)+\alpha_2 \mathcal{C}_k(t)=0 \label{eq:MCT}
\end{equation}
where, usually $\mathcal{K}(k,t-\tau)\propto \mathcal{C}^2_k(t-\tau)$ and $\alpha_i=const$. Juxtaposing \eqref{eq:C} and \eqref{eq:MCT} one can instantly notice that both describe the motion of a damped oscillator. However our theory is over-damped (no $\ddot{\mathcal{C}}_{k}(t)$ term) and makes use of the 'trivial' memory kernel, $\rho \hat U_k\delta(t-\tau)/\Lambda_{\Gamma,k}$.

Quick inspection of \eqref{eq:C} shows that our dynamics cannot describe the two-step relaxation, yet it still predicts the ergodic and non-ergodic phase. \eqref{eq:C} is satisfied by $\mathcal{C}_k(t)\propto \exp\left(-\rho \hat U_k t/\Lambda_{\Gamma,k} \right)$, which either decays exponentially fast to 0 or, for $\Lambda_{\Gamma,k}\to+\infty$, behaves as $\mathcal{C}_k(t)=const.$, which reproduces the saturation in the arrested state. Thus, apparently, our theory distils the simplified two-phase behaviour from MCT. The lack of the $\beta$-relaxation in the SCN-based model is no surprise, as the memory effects are neglected therein. This corresponds to the looking at the longer time-scale. However the presence of the non-ergodic phase suggests an intriguing conclusion: in the ideal glass transition the $\alpha$-relaxation (and so the arrest effect) is mostly governed by the spatial correlations. Although the memory effects and temporal correlations also affect this regime to a certain extent, the spatial aspect is apparently the backbone of the transition. 

This is a novel insight, as MCT usually treats the spatial inhomogeneity implicitly, putting more emphasis on the temporal evolution. One exception is e.g. the work of Reichman and Miyazaki \cite{bib:miyazaki}, who constructed the non-equilibrium (though based on the linear Onsager relations) density field formulation of MCT. Interestingly, in their Langevin equation for the evolution of the density, the fluctuation-dissipation relation explicitly leads to SCN. This is the manifestation of the local dependence between the density and the amplitude of noise, i.e. it introduces the fluctuating viscosity landscape. In our approach we achieve a similar effect, but by the means of the spatially variant inter-diffusivities. Thus, our viscosity landscape is dynamic and evolves with the change in the positions of particles. Yet, the theory of Reichman and Miyazaki eventually leads to the MCT-type equation in the generic form of \eqref{eq:MCT}, with only a slightly modified memory kernel. While being more realistic, this form still obscures the role of SCN. 

Summarizing this comparison, the similarities between MCT and the SCN-induced arrest are remarkable. Thus, we conclude that the transition embedded in our model is, in fact, a simplified variant of the ideal glass transition described by MCT. As such, our model provides a certain approximation for the dynamics in the supercooled state, but is not able to describe the glassy dynamics itself, in a similar manner as classic MCT fails to do so. However, our model can be studied analytically to a much greater degree than MCT, providing a complete insight into the mechanism of the SCN-induced arrest. Although stemming from the similar considerations, the SCN-based model employs a different set of assumptions and approximations, arriving at results that seem complementary to the classic MCT. 

\section*{Discussion in the context of glass research}\label{sec:discussion}
From the preceding comparison it is clear that our theory is closely related to MCT. However, MCT is well-known for its deficiencies \cite{bib:rev1, bib:mct_bad}, as the ideal glass transition, which MCT predicts, is not observed in experiments. One reason for this is that the glass transition seems to involve an intrinsically non-equilibrium component \cite{bib:non_eq}. Certain modern extensions aim at incorporating this aspect  into MCT \cite{bib:szamel, bib:feng}. Our theory, as presented in this work, is fully consistent only in the equilibrium conditions, i.e. as we derive the FDR with the help of the Boltzmann distribution, one should also calculate the noise correlation matrix $\textbf{H}$ using the same distribution. However, the Langevin dynamics it is not restricted to the equilibrium regime, i.e. one can start such dynamics also from the non-equilibrium initial conditions. In this case, it describes the equilibration process. Such results approximate the actual evolution of the system in an uncontrollable manner, but they must converge to the exact behaviour in the vicinity of equilibrium. The accuracy of our model in the non-equilibrium regime could be improved in two ways. A simpler option is to employ a non-equilibrium correlation function $H(r,t)$ in the current version of our FDR and assume its adiabatically slow evolution. This would result in a time-dependent $\pmb{\Gamma}$. The other approach is to derive our theory, but using a non-equilibrium distribution instead of Boltzmann $P_s$, if it can be postulated in a particular case. However, both require a certain additional input from non-equilibrium considerations, which is beyond the scope of this paper. 

Another important branch of the contemporary glass research is focused on the higher, three- and four-point correlation functions, which are well known to capture the dynamic characteristics of glasses \cite{bib:rev1,bib:4point}. The direct comparison of our theory with these results would require us to calculate the correlators of particle displacements. However, this could be done, provided that the explicit trajectories $x_i(t)$ are known. This is a formidable task since the spatially variant $\Gamma(x_i-x_j)$ makes the equations highly non-linear, even if the interactions are simple, e.g. harmonic. In the same context, it is also true that the assumption of the Gaussian SCN is insufficient, as it introduces only the two-point correlations at the microscopic level. Conversely, it has been already shown with the exact MCT involving the multi-point correlations\cite{bib:exact_mct} that the non-Gaussian fluctuations also contribute to the system behaviour, though this theory is challenging to apply. However, this suggest that our results could be improved by utilizing the non-Gaussian SCN (e.g. the heavy-tailed stochastic processes). Nevertheless, there is a correspondence between the multi-point correlation functions and the collective dissipation length. The former allows us to find how many neighbouring particles are involved in the collective rearrangements (i.e. the cooperation number \cite{bib:coop}), which in the three-dimensional systems is $\simeq10-20$. The collective dissipation length $\tau$ provides a similar estimate, indicating a rapid rise in the number of neighbours involved in the dissipation as the density grows. 

From the experimental perspective, the dependence of our theory on the noise correlation function $H(r)$ and $\lambda$ is inconvenient, as these magnitudes are difficult to measure. When a colloidal binary mixtures is suitable for the confocal microscopy, $\lambda$ can be obtained from the direct tracking of particles, though it requires that the system is already well-within the regime of the slowed-down dynamics\cite{bib:donth, bib:weeks}. A more direct measurements of the correlated noise acting on the larger particles could be accomplished with the use of e.g. optical tweezers \cite{bib:grier,bib:yodh}, though we are not aware of any such results. However, knowing the direct forces acting on a particle one could estimate the correlation of the effective forces, which can serve as a proxy for $H(r)$. 

The $\tilde \psi_c$ which we predict for different $\lambda$ and exponential correlations fits well into the known experimental and theoretical boundaries. The critical packing for vitrification is often related to the limit of random close packing and varies with dimensionality \cite{bib:ciesla}, type of particles, but also with the experimental or simulation protocol \cite{bib:rcp}. In particular, one should distinguish between Brownian glasses and glasses under the shear stress \cite{bib:jamming}. The reported values for mono-disperse colloids typically read 0.56-0.64 for the three dimensions \cite{bib:phi,bib:pusey}, 0.79-0.84 for two \cite{bib:dim,bib:2d,bib:2d_up} and approximately 1.0 for one. The model presented here is one-dimensional and Fig. \ref{fig:crit_psi}a and \ref{fig:hs2}a show that $\tilde \psi_c\simeq 1$ corresponds to the very short $\lambda$. The preliminary results for the higher-dimensional variant of our theory also indicate that $\tilde \psi_c$ in two and three dimensions can be reproduced for physically reasonable $\lambda$.

Experiments also provide an empirical dependence $\propto \exp(1.15\tilde \psi/(\tilde \psi_c-\tilde \psi))$ for viscosity \cite{bib:rev2,bib:phi}. Our model does not reproduce this shape, i.e. we predict the slower increase for most $\tilde \psi$, which becomes steep only in the vicinity of $\tilde \psi_c$. However, this empirical dependence comes from the shear viscosimetry, which involves external forces, not included in our model. The viscosity should also depend on how fast the control parameter is changed \cite{bib:ritland, bib:bartenev}, which is not reflected in our results. However, we use constant $\lambda$, while in the molecular systems it is likely to change with $\psi$. Thus, the differences might be partially remedied by the $\psi$-dependent noise correlation length. 

This also points to an important characteristic of our model, i.e. the effect of arrest is practically separated from the evolution of $\lambda$ and $H(r)$.  These magnitudes enter our FDR in self-contained manner, acting as the parameters for the mechanism of arrest. This is also reflected by the difference between $\lambda$ and the characteristic length of collective dissipation $\tau$. The latter comes from FDR and might become divergent independently from $\lambda$, which remains finite. While $\lambda$ is expected to increase with $\psi$, the collective dissipation apparently does not depend on the microscopic details of this process. Thus SCN acts as an intermediate \emph{cause} for the transition. This introduces a new perspective on the current advances in the determination of the characteristic length-scale for glass transition\cite{bib:length}. 

Although the SCN-induced arrest is only a simplified variant of the ideal glass transition and as such, it fails to describe some major aspects of the actual physical phenomenon, it brings in several novel ideas that might become useful in glass-related research. The fact that pure SCN can solely arrest the dynamics of a disordered system is a new and remarkable observation. It suggests another path to build glass transition theories, i.e. to start from the SCN-like approximations as a basis and include the temporal evolution and non-equilibrium effects as higher-order corrections. Whether this leads to a better analytical insight into the glass transition should be investigated in the future. The concept of the divergent collective dissipation length is also of special interest, as the profound change in dynamics without a corresponding change in structure is an important characteristics of the glass transition. A physical quantity capable of capturing this type of behaviour, analogous to one derived for SCN, would be of particular interest in the glass mechanics. Finally, in the grand perspective, our model shows how thermodynamic state of a system can be reflected in the effective dynamics via the properties of noise and dissipation. The FDR with embedded phase transitions is a valuable tool in the simulations of complex, multicomponent systems as it is computationally cheaper than full-scale molecular dynamics simulations. For this reason it can be employed as a component of some higher-level modelling. In the future, extending this approach to include other thermodynamic transitions, e.g. crystallization, would be a desirable achievement. 

\section*{Conclusions}
In this work we have identified SCN as the factor capable of inducing and controlling the critical slow-down of dynamics in disordered molecular systems at finite temperatures. The mechanism of this molecular arrest is analytically explained, i.e. the spectrum of the friction-response matrix is shown to act as the order parameter and the collective dissipation length is shown to diverge at the transition point. The theory is also identified as the simplified variant of MCT. The model might be seen as an effective, one component description of a binary mixture, reproducing a majority of disordered phases encountered in such systems. Our results contribute to the understanding of the role played by the spatial correlations in the physics of molecular arrest. They also suggest several further questions regarding the influence of such factors as: dimensionality, exact correlation functions, different types of interactions and non-equilibrium effects, which should be the matter of the future investigations.

\section*{Acknowledgments}
The authors gratefully acknowledge the National Science Center, Poland for the grant support (2014/13/B/ST2/02014).

\section*{Author contributions statement}
M. M.: obtained the main analytical and numerical results and wrote the manuscript, P.F.G.: general discussions and manuscript review

\section*{Competing interests}
The authors declare no competing interests. 

\numberwithin{equation}{subsection}
\section*{Appendices}
\renewcommand\thesubsection{\Alph{subsection}}
\subsection{Stochastic orthogonality} \label{app:stochastic}
We will justify the equality $Q^TQ=\textbf{1}_M$ in greater detail. A single element of this matrix product reads:
\begin{equation}
\begin{split}
&[Q^TQ]_{kk'}=\sum_i^N Q_{ik}Q_{ik'}=\frac{1}{N}\sum_i^N \left( \cos \frac{2\pi (k-k')x_i}{L}+\sin\frac{2\pi (k+k')x_i}{L} \right)
\end{split}
\end{equation}
where we have used the trigonometric identities. Assuming that $x_i$ has the homogeneous distribution $p(x_i)=L^{-1}$, one can calculate that the variance of $s_{ki}=\sin \frac{2\pi k x_i}{L}$ reads:
\begin{equation}
\begin{split}
<s_{ik}^2>&=L^{-1}\iint_{-L/2}^{L/2}dx_ids_i s_{ik}^2 \delta\left(s_{ik}-\sin \frac{2\pi k x_i}{L}\right)=\frac{1}{2}(1-\delta_{k0})
\end{split}
\end{equation}
Similar result is obtained for $c_{ik}=\cos \frac{2\pi k x_i}{L}$, i.e.:
\begin{equation}
<c_{ik}^2>=\frac{1}{2}(1-\delta_{k0})+\delta_{k0}
\end{equation}
This shows that the variances of $s_{ik}$ and $c_{ik}$ are finite. What follows, the variables:
\begin{align}
S_{k\neq 0}=\frac{1}{N}\sum_i^Ns_{ik}&& C_{k\ne 0}=\frac{1}{N}\sum_i^Nc_{ik}
\end{align}
must come from the Gaussian distributions with the mean 0 and the variance equal $(2N)^{-1}$ (for large $N$), as guaranteed by the central limit theorem. Therefore, the variable:
\begin{equation}
[Q^TQ]_{kk'}=C_{k-k'}+S_{k+k'}
\end{equation}
is the sum of two Gaussian variables. This means that $[Q^TQ]_{kk'}$ is also a Gaussian variable, with 0 mean and variance $N^{-1}$. This variance reduces to 0 as $N\to+\infty$. Thus, in the very large system we have:
\begin{equation}
\lim_{N\to+\infty}[Q^TQ]_{kk'}=\delta_{kk'}
\end{equation}
i.e. except for $k=k'$, the trigonometric functions statistically compensate to 0. 

\subsection{SCN and different interpretations of stochastic integrals}\label{app:stoch_int}
First, we will identify the correction terms for the general choice of interpretation. This is a generalization of the considerations from the Ref. \cite{bib:lau} (especially equation 2.13 therein) to the multidimensional case. We consider a stochastic differential equation of the form:
\begin{equation}
\dot x_i=f_i(\vec{x})+\sum_k g_{ik}(\vec{x})\eta_k(t)
\end{equation}
and its short-time integrated version:
\begin{equation}
\begin{split}
&\Delta x_i(t)=\int_t^{t+\Delta t} ds \dot x_i=f_i(\vec{x}(t))\Delta t+\sum_j\int_{t}^{t+\Delta t}ds g_{ik}(\vec{x}(s))\eta_k(s)
\end{split}
\end{equation}
The integral over the stochastic part can be specified as:
\begin{equation}
\begin{split}
I_\alpha&=\int_{t}^{t+\Delta t}ds g_{ik}(\vec{x}(x))\eta_k(s)=g_{ik}(\alpha \vec{x}(t+\Delta t)+(1-\alpha)\vec{x}(t))\int_{t}^{t+\delta t}ds \eta_k(s)
\end{split}
\end{equation}
where the choice of $\alpha$ corresponds to some interpretation (e.g. $\alpha=0$ for Ito, $\alpha=1/2$ for Stratonovich, but other values are also allowed). Following the steps taken in the Ref. \cite{bib:lau}, one can calculate:
\begin{equation}
<\Delta x_i>=f_i(\vec{x}(t))\Delta t+\alpha \sum_{j,k} g_{kj}(\vec{x}(t))\partial_{x_j} g_{ik}(\vec{x}(t))
\end{equation}
The $\alpha$-dependent term is the additional drift related to the noise interpretation, i.e.:
\begin{equation}
C_i=\alpha \sum_{j,k} g_{kj}(\vec{x}(t))\partial_{x_j} g_{ik}(\vec{x}(t))
\end{equation}
We can now calculate $\vec{C}$ for equation \eqref{eq:Lan1.5}:
\begin{equation}
\begin{split}
&C_i/\sigma^2=\alpha\sum_j^N \sum_{k=-(M+1)/2}^{k=(M+1)/2} Q_{jk}\Lambda_{\Gamma,k}^{-1}\Lambda_{H,k}^{1/2} \partial_{x_j}\left(Q_{ik}\Lambda_{\Gamma,k}^{-1}\Lambda_{H,k}^{1/2} \right)=\alpha\sum_{k=-(M+1)/2}^{k=(M+1)/2} \frac{2\pi k}{L}Q_{ik}Q_{i,-k}\Lambda_{\Gamma,k}^{-2}\Lambda_{H,k}
\end{split}
\end{equation}
where we have used $\partial_{x_j}Q_{ik}=\delta_{ij}\frac{2\pi k}{L}Q_{i,-k}$. One can realize now that since $\frac{2\pi k}{L}Q_{ik}Q_{i,-k}\Lambda_{\Gamma,k}^{-2}\Lambda_{H,k}$ is antisymmetric in $k$, we obtain: 
\begin{equation}
C_i=\alpha\sigma^2\sum_{k=-(M+1)/2}^{k=(M+1)/2} \frac{2\pi k}{L}Q_{ik}Q_{i,-k}\Lambda_{\Gamma,k}^{-2}\Lambda_{H,k}=0
\end{equation}
This means that for SCN every interpretation is equivalent as they produce no additional drift.

\subsection{Derivation of $\Lambda_{\Gamma}$ from the Fokker-Planck equation} \label{app:FP}
In this appendix we will determine the dissipation function $\Gamma(r)$ using the stationary Fokker-Planck \eqref{eq:FPE0} with the assumption that $P_s$ is the Boltzmann distribution. In the first step, we will separate the contributions from the correlated and non-correlated dynamics. We expect that as a certain correlation length $\lambda$ falls to zero, $H(r)\to \frac{a_H}{\rho} \delta(r)$. Thus, $\textbf{H}$ can be decomposed as:
\begin{equation}
\textbf{H}=Q\Lambda_H Q^T= Q(a_H\textbf{1}_M+\Lambda_{\Delta H})Q^T
\end{equation}
where $a_H\textbf{1}_M$ corresponds to the Dirac-delta correlations in the $\lambda\to0$ limit and $\Lambda_{\Delta H}$ describes the finite-range correlations. Similarly, we can represent $\pmb{\Gamma}$ and its inverse:
\begin{align}
\pmb{\Gamma}=\gamma a_\Gamma Q(\textbf{1}_M+\Lambda_{\Delta \Gamma})Q^T \label{eq:Gamma} &&\\
\gamma a_\Gamma \pmb{\Gamma}^{-1}=Q(\textbf{1}_M-\Lambda_K)Q^T=(\textbf{1}_N-\textbf{K})&&\Lambda_{\Delta \Gamma,k}=\frac{\Lambda_{K,k}}{1-\Lambda_{K,k}} \label{eq:Gamma2}
\end{align}
where $\Lambda_{\Delta \Gamma}$ and $\textbf{K}=Q\Lambda_KQ^T$ carry the information on how the finite-range correlations influence $\pmb{\Gamma}$ and $\pmb{\Gamma}^{-1}$, $\gamma$ is the usual hydrodynamic friction and $a_\Gamma$ will be later related to $a_H$. The equation \eqref{eq:Lan1.5} now reads:
\begin{equation}
\begin{split}
 \dot{\vec{x}}=&\frac{1}{\gamma a_\Gamma}Q(\textbf{1}_M-\Lambda_{K})Q^T\vec{F}+ \frac{\sigma}{\gamma a_\Gamma} Q (\textbf{1}_M-\Lambda_K)\sqrt{a_H\textbf{1}_N+\Lambda_{\Delta H}}\vec{\eta} \label{eq:Lan2}
 \end{split}
\end{equation}
Let us define the auxiliary diffusion matrix $\textbf{D}$, such that:
\begin{align}
Q\Lambda_\Gamma^{-1}\Lambda_H\Lambda_\Gamma^{-1}Q^T=a_H\textbf{1}_N+\textbf{D}&&
\textbf{D}=Q\Lambda_DQ^T=Q(\Lambda_{\Delta H}-2\Lambda_{H}\Lambda_K+\Lambda_H\Lambda_K^2)Q^T \label{eq:aux2}
\end{align}
and $\Lambda_D$ is the spectrum of $\textbf{D}$. With the aid of $\textbf{D}$ we can now rewrite down the stationary Fokker-Planck equation as:
\begin{equation}
\begin{split}
&0=\frac{1}{\gamma a_\Gamma}\sum_i^N\partial_{x_i}\left( F_iP_s-\frac{\sigma^2a_H}{2\gamma a_{\Gamma}}\partial_{x_i}P_s\right)-\frac{1}{\gamma a_\Gamma}\sum_i^N\partial_{x_i}\sum_j^N\left(\textbf{K}_{ij}F_jP_s+\frac{\sigma^2}{2\gamma a_{\Gamma}}\partial_{x_j}(\textbf{D}_{ij}P_s)\right)
\end{split} \label{eq:FP}
\end{equation}
In the limit of the non-correlated noise we expect both $\Lambda_{\Delta H}\to0$ and $\Lambda_K\to 0$, so \eqref{eq:FP} reduces to the case of the ordinary diffusion. In this case, remembering that $\partial_{x_j}P_s=\beta F_j P_s$, the equation \eqref{eq:FP} is satisfied if $\beta \sigma^2a_H/(2\gamma a_\Gamma)=1$. This means that the classical dissipation relation $\beta \sigma^2/(2\gamma)=1$ holds and $a_H=a_\Gamma$, i.e. we should use the same representation for the diagonal part of $\pmb{\Gamma}$ and $\textbf{H}$. 

After employing all these identities, the first sum in \eqref{eq:FP} becomes identically equal to 0. We can take now the derivative over $x_j$ in the remaining part of \eqref{eq:FP} to obtain:
\begin{equation}
\begin{split}
&0=-\sum_i^N\partial_{x_i}\left[\frac{1}{\gamma a_H^2}P_s \sum_j^N\left(a_H \textbf{K}_{ij}F_j+\frac{\sigma^2}{2\gamma }\partial_{x_j} \textbf{D}_{ij}+\frac{\beta \sigma^2}{2\gamma } \textbf{D}_{ij}F_j)\right) \right]
\end{split}\label{eq:FPaux}
\end{equation}
We will now represent all the terms under sum with the aid of the $Q_{ik}$ functions:
\begin{gather}
\begin{split}
F(x_i-x_j)&=\frac{1}{L} \sum_{k=-(M-1)/2}^{(M-1)/2} \hat F_k \sin \frac{2\pi k (x_i-x_j)}{L}=\rho\sum_{k=-(M-1)/2}^{(M-1)/2} \hat F_k Q_{ik}Q_{j,-k} \label{eq:F_exp}
\end{split}\\
\begin{split}
\textbf{K}_{ij}&=\sum_{k=-(M-1)/2}^{(M-1)/L}\Lambda_{K,k}Q_{ik}Q_{jk}=\frac{1}{N}\sum_{k=-(M-1)/2}^{(M-1)/L}\Lambda_{K,k} \cos \frac{2\pi k (x_i-x_j)}{L}
\end{split}\\
\begin{split}
\textbf{D}_{ij}&=\sum_{k=-(M-1)/2}^{(M-1)/L}\Lambda_{D,k}Q_{ik}Q_{jk}=\frac{1}{N}\sum_{k=-(M-1)/2}^{(M-1)/L}\Lambda_{D,k} \cos \frac{2\pi k (x_i-x_j)}{L}
\end{split}\label{eq:Dexpand}
\end{gather}
where we frequently use the cancellation of the antisymmetric terms under the summation over $k$. We can further use these expansions to calculate:
\begin{equation}
\begin{split}
&\sum_j^N \textbf{K}_{ij} F_j=\rho \sum_{k,k'=-(M-1)/2}^{(M-1)/2}\sum_{j,l}^N \hat F_{k'}\Lambda_{K,k}Q_{ik} Q_{jk} Q_{jk'}Q_{l,-k'}
\end{split}
\end{equation}
By the stochastic orthogonality, $\sum_j^N Q_{jk}Q_{jk'}\simeq\delta_{kk'}$, this further reads:
\begin{equation}
\begin{split}
&\sum_{j}^N \textbf{K}_{ij} F_{j}\simeq \rho\sum_{k=-(M-1)/2}^{(M-1)/2}\sum_{l}^N \hat F_k \Lambda_{K,k} Q_{ik} Q_{l,-k}=\frac{\rho }{N}\sum_l^N\sum_{k=-(M-1)/2}^{(M-1)/2}\hat F_k\Lambda_{K,k} \sin \frac{2\pi k(x_i-x_l)}{L}
\end{split}
\end{equation}
In exactly the same way we calculate:
\begin{equation}
\begin{split}
&\sum_j^N \textbf{D}_{ij}F_j=\rho\sum_{k,k'=-(M-1)/2}^{(M-1)/2} \sum_{j,l}^N\Lambda_{D,k}\hat F_k Q_{ik}Q_{jk}Q_{jk'}Q_{l,-k'}=\frac{\rho }{N}\sum_l^N\sum_{k=-(M-1)/2}^{(M-1)/2}\Lambda_{D,k}\hat F_k\sin \frac{2\pi k(x_i-x_l)}{L}
\end{split}
\end{equation}
Eventually, the last term reads:
\begin{equation}
\sum_j^N\partial_{x_j}\textbf{D}_{ij}=\frac{1}{N}\sum_{k=-(M-1)/2}^{(M-1)/2}\frac{2\pi k}{L}\Lambda_{D,k}\sin \frac{2\pi k(x_i-x_j)}{L}
\end{equation}
Inserting these results into \eqref{eq:FPaux}, we obtain:
\begin{equation}
\begin{split}
&0=-\sum_i^N\partial_{x_i}\left[\frac{P_s}{\gamma a_H L }\sum_j^N\sum_{k=-(M-1)/2}^{(M-1)/2}\sin \frac{2\pi k (x_i-x_j)}{L}\left( a_H \Lambda_{K,k}\hat F_{k}+\frac{\sigma^2}{2\gamma\rho}\frac{2\pi k}{L}\Lambda_{D,k}+\frac{\beta \sigma^2}{2\gamma}\Lambda_{D,k}\hat F_k\right) \right]
\end{split}
\end{equation}
Finally, one can realize that since $F(r)=-\partial_r U(r)$, then $\hat F_k=\frac{2\pi k}{L}\hat U_k$. Using this representation and the dissipation relation $\frac{\beta \sigma^2}{2\gamma}=1$ we can factor out all the common terms, so \eqref{eq:FP2} becomes:
\begin{equation}
\begin{split}
0=&\sum_{i,j}^N \partial_{x_i}\left\{ \sum_{k=-(M-1)/2}^{(M-1)/2}\frac{2\pi k}{L} \sin \frac{2\pi k (x_i-x_j)}{L} \frac{P_s}{\gamma  a_H^2 L} \left[a_H \hat U_k\Lambda_{K,k}+\left(\frac{1}{\beta\rho}+\hat U_k\right)\Lambda_{D,k} \right] \right\}
\end{split}\label{eq:FP2}
\end{equation}
Inserting the definition of $\Lambda_D$ \eqref{eq:aux2} and rearranging, we obtain the quadratic equation for a single mode $\Lambda_{K,k}$:
\begin{align}
0=\Lambda_{\Delta H,k}+\left( a_H\Phi_k-2\Lambda_{H,k}\right)\Lambda_{K,k}+\Lambda_{H,k}\Lambda_{K,k}^2 &&\Phi_k= \beta\rho \hat U_k/(1+\beta\rho \hat U_k)\label{eq:quadratic}
\end{align}
 Before we solve this equation, let us assume that $\Lambda_{\Delta H,k}/a_H\le 1$, i.e. the Fourier spectrum of the correlation function is limited by the spectrum of the delta-like correlations. From the fact that $\Lambda_{\Delta H,k}=\Lambda_{H,k}-a_H$ and $\Lambda_{H,k}\le a_H$ it follows that $\Lambda_{\Delta H,k}<0$. This means that \eqref{eq:quadratic} has two solutions as its determinant is non-negative:
\begin{align}
\Lambda_{K,k}^{(\pm)}=\frac{-(a_H \Phi_k-2\Lambda_{H,k})\pm\sqrt{\det_k}}{2\Lambda_{H,k}}&&{\det}_k=(a_H\Phi_k-2\Lambda_{H,k})^2-4\Lambda_{H,k}\Lambda_{\Delta H,k}
\end{align}
However, we will reduce the number of solutions by demanding that in the non-correlated case ($\Lambda_{\Delta H,k}\to a_H$) it is true that $\Lambda_{K,k}^{(\pm)}\to0$. In order to impose this restriction, we can rewrite $\det_k$ in the following way:
\begin{equation}
{\det}_k=a_H^2\left( \Phi_k-2\right)^2-4a_H\left(\Phi_k-2\right)\Lambda_{\Delta H,k}-4a_H\Lambda_{\Delta H,k}=a_H^2(\Phi_k-2)^2\left(1-\left(\frac{4}{\Phi_k-2}+\frac{4}{(\Phi_k-2)^2}\right)\frac{\Lambda_{\Delta H,k}}{a_H}\right)
\end{equation}
As $|\Lambda_{\Delta H}/a_H|\le 1$, we can use it as an expansion parameter for $\sqrt{\det_k}$, thus:
\begin{equation}
\Lambda_{K,k}^{(\pm)}\simeq\frac{-(a_H\Phi_k-2\Lambda_{H,k})}{2\Lambda_{H,k}}\pm \frac{a_H|\Phi_k-2|}{2\Lambda_{H,k}} \left(1-\left(\frac{2}{\Phi_k-2}+\frac{2}{(\Phi_k-2)^2}\right)\frac{\Lambda_{\Delta H,k}}{a_H}\right)
\end{equation}
In the limit $\Lambda_{H,k}\to a_H$, this reduces to $\Lambda_{K,k}^{(\pm)}=(-(\Phi_k-2)\pm|\Phi_k-2|)/2$. This shows that the condition $\Lambda_{K,k}=0$ is satisfied only interval-wise for either solution, i.e. by $\Lambda_{K,k}^{(+)}$ for $\Phi_k>2$ and by $\Lambda_{K,k}^{(-)}$ for $\Phi_k<2$. Thus, we construct the appropriate solution from these two functions on the complementary intervals. After reducing all the cancelling terms in the each interval, the eventual result proves to be a continuous function:
\begin{equation}
\Lambda_{K,k}=-\frac{\Lambda_{\Delta H,k}}{\Lambda_{H,k}(\Phi_k-2)} \label{eq:sol}
\end{equation}
Let us inserted this formula in \eqref{eq:Gamma2} and \eqref{eq:Gamma} and use the definition of $\Phi_k$, then we obtain:
\begin{gather}
\Lambda_{\Gamma,k}=\gamma a_\Gamma\frac{\Lambda_{H,k}(2+\beta \rho \hat U_k)}{\Lambda_{H,k}+a_H(1+\beta \rho \hat U_k)}\\
\pmb{\Gamma}_{ij}=\sum_{k=-(M-1)/2}^{(M-1)/2}\Lambda_{\Gamma,k}Q_{ik}Q_{jk}=\frac{1}{N}\sum_{k=-(M-1)/2}^{(M-1)/2}\Lambda_{\Gamma,k}\cos\left(\frac{2\pi k}{L}(x_i-x_j)\right)
\end{gather}
We can now use the identity $a_\Gamma=a_H$ and define $\hat h_k=\rho \hat H_k/a_H$. As a final step, the thermodynamic limit is applied, i.e. $N\to+\infty$, $L\to +\infty$ (while $N/L=\rho$) and $\sum_{k=-(M-1)/2}^{(M-1)/2}\to\frac{L}{2\pi}\int_{-\pi m/d}^{\pi m/d}dk$ as $\frac{2\pi k}{L}\to k$. $d$ is the particle diameter, which is introduced here to make the continuous cut-off parameter $m$ dimensionless. Eventually we obtain the main result of this paper, formula \eqref{eq:Gamma_fin}

\subsection{Mean behaviour of $\pmb{\Gamma}$, the $\pmb{\Gamma}\propto(\psi_c-\psi)^{-1/2}$ dependence and the approximation of $\pmb{\Gamma}_{ij}$} \label{app:exponent}
We can calculate the mean value of each $\pmb{\Gamma}_{ij}$ in a homogeneous system, where the probability of finding a particle reads $p(x_i)=L^{-1}$. Let us denote $<f(x_n)>_x=\prod_{i=1}^N\int_{-L/2}^{L/2}dx_ip(x_i)f(x_n)$. Since $<\cos k(x_i-x_j)>_x=\delta (k)$, we obtain:
\begin{align}
<\pmb{\Gamma}_{ii}>_x=\pmb{\Gamma}_{ii}=\Gamma(0)&&<\pmb{\Gamma}_{i\neq j}>_x=\frac{\gamma a_H}{2\pi \rho}\frac{\hat h(0)[2+\beta \rho \hat U(0)]}{1+\hat h(0)+\beta \rho \hat U(0)} \label{eq:mean_gamma}
\end{align}
When $H(r)$ is normalized in such a manner that $ \hat h(0)=1$, $<\pmb{\Gamma}_{i\neq j}>_x$ reduces to $\frac{\gamma a_H}{2\pi \rho}$, i.e. the constant prefactor of $\Gamma(0)$. Thus, the ratio $<\pmb{\Gamma}_{i\neq j}>_x/<\pmb{\Gamma}_{ii}>_x$ is proportional to $1/\pmb{\Gamma}_{ii}$. Since the diagonal terms $\pmb{\Gamma}_{ii}$ can grow in an unbounded manner, thus the averaged-out matrix $<\pmb{\Gamma}>$ can be easily dominated by its diagonal terms.

In the vicinity of critical packing $\psi\simeq \psi_c $, the integral \eqref{eq:Gamma_fin} expressing any $\pmb{\Gamma}_{ij}$ is dominated by the contribution from the singularity at $k_0$. Therefore, we can approximate the integrand by its value at this point. Expanding $f(k,\psi)$ given by \eqref{eq:cond} around $k_0$, up to the second order, we get:
\begin{equation}
\begin{split}
f(k,\psi)&\simeq f(k_0,\psi)+f''(k_0,\psi)\frac{(k-k_0)^2}{2}=f(k_0,\psi_c)+f(k_0,\psi)-f(k_0,\psi_c)+f''(k_0,\psi)\frac{(k-k_0)^2}{2}=\\
&=f(k_0,\psi_c)+(\psi_c-\psi)|\hat U(k_0)|+f''(k_0,\psi)\frac{(k-k_0)^2}{2}
\end{split}
\end{equation}
One must also remember that $f(k_0,\psi_c)=0$, by definition. The above expansion approximates the denominator of the integrand. We can also treat the nominator as constant, thus:
\begin{equation}
\begin{split}
\frac{\pmb{\Gamma}_{ij}(r)}{\gamma}\simeq &\frac{1}{\pi}\int_0^{\frac{\pi m}{d}}dk\frac{\hat h(k_0)(2+\psi \hat U(k_0))\cos k_0 r}{(\psi_c-\psi)|\hat U(k_0)|+f''(k_0,\psi)\frac{(k-k_0)^2}{2}}=\\
&=\frac{1}{\sqrt{\psi_c-\psi}}\frac{\hat h(k_0)(2+\psi \hat U(k_0))\cos k_0r}{\sqrt{|\hat U(k_0)|f''(k_0,\psi)/2}}\left.\arctan \left(\sqrt{\frac{f''(k_0,\psi)}{2|\hat U(k_0)|}}(k-k_0) \right) \right|_{k=0}^{k=\frac{m\pi}{d}}
\end{split}\label{eq:app_gamma1}
\end{equation}
This shows that the dominant behaviour in $\psi$ is of the $\propto (\psi_c-\psi)^{-1/2}$ type. We can also identify that $\pmb{\Gamma}_{ij}=\pmb{\Gamma}_{ii}\cos k_0 r$, since $\pmb{\Gamma}_{ii}=\Gamma(0)$. However, this approximation of the spatial dependencies in $\Gamma(r)$ works only very close to $\psi_c$.

If we can effectively treat the cut-off as very large, i.e. $\frac{\pi m}{d}\to +\infty$, a more accurate approach is possible. We assume now that the integrand is dominated by the singularity at $k_0$, but the oscillations are still important, i.e. we approximate:
\begin{equation}
\begin{split}
\frac{\pmb{\Gamma}_{ij}(r)}{\gamma}\simeq &\frac{1}{2\pi}\int_{-\infty}^{+\infty}dk\frac{\hat h(k_0)(2+\psi \hat U(k_0))\cos k r}{(\psi_c-\psi)|\hat U(k_0)|+f''(k_0,\psi)\frac{(k-k_0)^2}{2}}
\end{split}
\end{equation}
As we change the integration variables to $q=k-k_0$, we expand $\cos kr=\cos k_0r \cos qr-\sin k_0r \sin qr$. The latter term disappears under the integral due to its antisymmetric character. Therefore, we are left with:
\begin{equation}
\begin{split}
\frac{\pmb{\Gamma}_{ij}(r)}{\gamma}\simeq &\frac{\cos k_0 r}{2\pi}\int_{-\infty}^{+\infty}dq\frac{\hat h(k_0)(2+\psi \hat U(k_0))\cos q r}{(\psi_c-\psi)|\hat U(k_0)|+f''(k_0,\psi)\frac{q^2}{2}}=\\
&=\frac{\hat h(k_0)(2+\psi \hat U(k_0)) }{\sqrt{2(\psi_c-\psi)f''(k_0,\psi)|\hat U(k_0)|}}\cos k_0 r \exp\left( -\frac{|r|}{\sqrt{\frac{f''(k_0,\psi)}{2(\psi_c-\psi)|\hat U(k_0)|}}}\right)
\end{split}
\end{equation}
The formula \eqref{eq:non_diag_gamma} follows from this result.

\subsection{Derivation for the HS case}\label{app:HS}
In order to solve the equation \eqref{eq:hs1} we can act on its both sides with $\int_{-\infty}^{+\infty}dr e^{ikr}$ to obtain:
\begin{equation}
-2 i\sin kd = \int_{-\infty}^{+\infty} dk'\left( \delta(k+k')-\frac{\sin(k+k')d}{\pi(k+k')}\right)f(k')
\end{equation}
We postulate now that $\hat f(k')=\lambda \sin k'd$ and it is enough to find $\lambda$. Thus we want to solve the problem:
\begin{equation}
-2i\sin kd =-\lambda\sin kd-\lambda\int_{-\infty}^{+\infty}dk' \frac{\sin(k+k')d\sin k'd}{\pi(k+k')}\label{eq:app_hs}
\end{equation}
The integral can be calculated exactly if we switch to the exponential representation and apply the residue theorem to the pole on the real axis, i.e.:
\begin{equation}
\begin{split}
&\int_{-\infty}^{+\infty}dk' \frac{\sin(k+k')d\sin k'd}{\pi(k+k')}=-\int_{-\infty}^{+\infty}dk'\frac{e^{i(k+2k')d}+e^{-i(k+2k')d}-2\cos kd}{4\pi(k+k')}=\\
&=-\frac{i}{4}(e^{-ikd}-e^{ikd})+0=-\frac{1}{2}\sin kd
\end{split}
\end{equation}
Substituting this result into \eqref{eq:app_hs} we obtain:
\begin{equation}
-2i\sin kd = -\lambda \sin kd +\frac{\lambda}{2}\sin kd
\end{equation}
and eventually $\lambda=4i$. This shows that our postulated solution is correct and it can be utilized to determine $\hat U_{HS}(k)$.

\subsection{The relation to MCT}\label{app:mct}
As a first step, we want to change the variables from $x_i$ to  $y_k$, as defined by \eqref{eq:y_k}. In Appendix \ref{app:stoch_int} we have already shown that SCN is insensitive to the change of noise interpretation, thus we can then instantly switch to the Stratonovich interpretation to make the calculations straightforward. In order to change variables, let us multiply \eqref{eq:Lan} by $Q^T$ from the left:
\begin{equation}
\gamma a_{\Gamma} \Lambda_{\Gamma} Q^T\dot {\vec{x}}=\vec{F}+\sigma \Lambda_H^{1/2}\vec{\eta} \label{eq:aux_Lan}
\end{equation}
$y_k$ is the function of $x_i$, which is the function of time. Thus the $M$-dimensional vector of $y_k$ satisfies:
\begin{equation}
\dot{\vec{y}}=-\frac{2\pi k}{L}Q^T\dot{\vec{x}}
\end{equation}
We can also explicitly identify the element of $Q^T\vec{F}$, using the Fourier expansion \eqref{eq:F_exp} for $F(x_i-x_j)$ and the stochastic orthogonality:
\begin{equation}
\begin{split}
[Q^T\vec{F}]_k&=\rho\sum_{i,j}^NQ_{ik}\sum_{k'=-(M+1)/L}^{(M+1)/L}\hat F_{k'} Q_{ik'}Q_{j,-k'}=\rho\hat F_k\sum_j^NQ_{j,-k}=\rho \hat F_k y_k=\rho\frac{2\pi k}{L} \hat U_k y_k
\end{split}
\end{equation}
because, in fact $y_k=\sum_i Q_{i,-k}$. Inserting these results in \eqref{eq:aux_Lan} leads to the equation \eqref{eq:Lan_y}. 

In order to establish the relation with MCT and $\mathcal{C}_k(t)$, one can realize that:
\begin{equation}
\begin{aligned}
\frac{y_k+y_{-k}}{2}=\frac{1}{N}\sum_i^N\cos\frac{2\pi k x_i}{L}&&\frac{y_{-k}-y_k}{2}=\frac{1}{N}\sum_i^N\sin\frac{2\pi k x_i}{L}
\end{aligned}
\end{equation}
Therefore, using these identities to expand \eqref{eq:def_C}, we obtain:

\begin{equation}
\begin{split}
\mathcal{C}_k(t)=&\left<\frac{1}{N}\sum_{i,j}^{N}e^{i\frac{2\pi k}{L}(x_i(0)-x_j(t))}\right>=\\
&=\left<\frac{1}{N}\sum_{i}^{N}\left(\cos\frac{2\pi k}{L}x_i(0)+i\sin\frac{2\pi k}{L}x_i(0)\right)\sum_{i}^{N}\left(\cos\frac{2\pi k}{L}x_j(t)-i\sin\frac{2\pi k}{L}x_j(t)\right)\right>=\\
=&\frac{1}{4}<\left(y_k(0)+y_{-k}(0)\right)\left(y_k(t)+y_{-k}(t)\right)>+\frac{1}{4}<\left(y_{-k}(0)-y_{k}(0)\right)\left(y_{-k}(t)-y_{k}(t)\right)>+\\
&+\frac{i}{4}<\left(y_k(0)+y_{-k}(0)\right)\left(y_{-k}(t)-y_{k}(t)\right)>-\frac{i}{4}<\left(y_{-k}(0)-y_{k}(0)\right)\left(y_k(t)+y_{-k}(t)\right)>=\\
=&\frac{1}{2}\left(<y_k(0)y_{k}(t)>+<y_{-k}(0)y_{-k}(t)> \right)+\frac{i}{2}\left(<y_k(0)y_{-k}(t)>-<y_{-k}(0)y_{k}(t)> \right)
\end{split}
\end{equation}
The equation for $\dot{\mathcal{C}}_k(t)$ follows instantly:
\begin{equation}
\dot{\mathcal{C}}_k(t)=\frac{1}{2}\left(<y_k(0)\dot y_{k}(t)>+<y_{-k}(0)\dot y_{-k}(t)> \right)+\frac{i}{2}\left(<y_k(0)\dot y_{-k}(t)>-<y_{-k}(0)\dot y_{k}(t)> \right)
\end{equation}
Finally, replacing $\dot y_k(t)$ with \eqref{eq:Lan_y}, assuming that the averages containing the noise terms $\eta_k$ disappear and remembering that $\hat U_k=\hat U_{-k}$ and $\Lambda_{\Gamma,k}=\Lambda_{\Gamma,-k}$, we obtain \eqref{eq:C}.


\begin{thebibliography}{99}
\bibitem{bib:rev1} L. Berthier, G. Biroli, Theoretical perspective on the glass transition and amorphous materials, Rev. Mod. Phys., 83, 2, 587 (2011)
\bibitem{bib:rev2} G. L. Hunter, E. R. Weeks, The physics of the colloidal glass transition, Rep. Prog. Phys., 75, 6, 066501 (2012)
\bibitem{bib:likos} C. N. Likos, Effective interactions in soft condensed matter physics, Phys. Rep., 348, 4, 267 (2001)
\bibitem{bib:MZ1} R. Zwanzig, Nonlinear generalized Langevin equations, J. Stat. Phys. 9, 3, 215 (1973)
\bibitem{bib:MZ2} H. Mori, Transport, Collective Motion, and Brownian Motion, Prog. Theor. Exp. Phys., 33, 3, 423 (1965)
\bibitem{bib:SCKou} S. C. Kou, Stochastic modeling in nanoscale biophysics: subdiffusion within proteins, Ann. Appl. Stat., 2, 2, 501 (2008)
\bibitem{bib:Dhont} J.K.G. Dhont \emph{An introduction to dynamics of colloids} ed. D. M\''{o}bius and R. Miller, (Elsevier, 1996)
\bibitem{bib:Deutch} J. M. Deutch, I. Oppenheim, Molecular Theory of Brownian Motion for Several Particles, J. Phys. Chem., 54, 8, 3547 (1971)
\bibitem{bib:hydrodynamic} D. L. Ermak, J. A. McCammon, Brownian dynamics with hydrodynamic interactions, J. Chem. Phys., 69, 4, 1352 (1978)
\bibitem{bib:SCN1} M. Majka, P.F. G\'{o}ra, Thermodynamically consistent Langevin dynamics with spatially correlated noise predicting frictionless regime and transient attraction effect, Phys. Rev. E, 94, 4,  042110 (2016)
\bibitem{bib:SCN2} M. Majka, P.F. G\'{o}ra, Collectivity in diffusion of colloidal particles: from effective interactions to spatially correlated noise, J. Phys. A: Math. Theor., 50, 5, 054004  (2017)
\bibitem{bib:polymers} M. Majka, P.F. G\'{o}ra, Polymer unfolding and motion synchronization induced by spatially correlated noise, Phys. Rev. E, 86, 5, 051122 (2012)
\bibitem{bib:hs1} M. Tokuyama, I. Oppenheim, On the theory of concentrated hard-sphere suspensions, Physica A, 216, 1-2, 85 (1995)
\bibitem{bib:hs2} M. Tokuyama, I. Oppenheim, Physica, Dynamics of hard-sphere suspensions, Phys. Rev. E, 50, 1, R16 (1994)


\bibitem{bib:cell1} M. Guo, A. J. Ehrlicher, M. H. Jensen, M. Renz, J. R. Moore, R. D. Goldman, J. Lippincott-Schwartz, F. C. Mackintosh and D. A. Weitz, Probing the stochastic, motor-driven properties of the cytoplasm using force spectrum microscopy, Cell,158, 4, 822 (2014)
\bibitem{bib:cell2} R. Kapral and A. S. Mikhailov, Stirring a fluid at low Reynolds numbers: Hydrodynamic collective effects of active proteins in biological cells, Physica D, 318, 100 (2016)
\bibitem{bib:miyazaki} K. Miyazaki and D. R. Reichman, Mode-coupling theory and the fluctuation-dissipation theorem for nonlinear Langevin equations with multiplicative noise, J. Phys. A: Math. Gen., 38, 20, L343 (2005)
\bibitem{bib:majka} M. Majka, P.F. G\'{o}ra, Analytical theory of effective interactions in binary colloidal systems of soft particles, Phys. Rev. E, 90, 3, 032303 (2014)
\bibitem{bib:lekkerkerker} H. N. W. Lekkerkerker, R. Tuinier, Colloids and the Depletion Interaction (Springer, London, 2011)

\bibitem{bib:pusey} P. N. Pusey, W. van Megen, Observation of a glass transition in suspensions of spherical colloidal particles, Phys. Rev. Lett., 59, 18, 2083 (1987)
\bibitem{bib:phi} Z. Cheng, J. Zhu, P. M. Chaikin, S.-E. Phan, W. B. Russel, Nature of the divergence in low shear viscosity of colloidal hard-sphere dispersions, Phys. Rev. E, 65, 4, 041405 (2002)
\bibitem{bib:likos_diff} S. Gupta, J. Stellbrink, E. Zaccarelli, C. N. Likos, M. Camargo, P. Holmqvist, J. Allgaier, L. Willner, D. Richter, Validity of the Stokes-Einstein relation in soft colloids up to the glass transition, Phys. Rev. Lett., 115, 12, 128302 (2015)
\bibitem{bib:mass_sims} P. Sibani and C. Svaneborg, Dynamics of dense hard sphere colloidal systems: A numerical analysis, Phys. Rev. E, 99, 4, 042607 (2019)
\bibitem{bib:ritland} H.N. Ritland, Density phenomena in the transformation range of a borosilicate crown glass, J. Am. Ceram. Soc., 37, 8, 370 (1954)
\bibitem{bib:bartenev} G. Bartenev, On the relation between the glass transition temperature of silicate glass and rate of cooling or heating, Dokl. Akad. Nauk SSSR 76, 2, 227 (1951)
\bibitem{bib:non_eq} L. Berthier and J. Kurchan, Non-equilibrium glass transitions in driven and active matter, Nat. Phys., 9, 5, 310 (2013)
\bibitem{bib:subdiff1} E. R. Weeks, D. A. Weitz, Subdiffusion and the cage effect studied near the colloidal glass transition, Chem. Phys., 284,1-2, 361 (2002)
\bibitem{bib:subdiff2} S. Boettcher, P. Sibani, Ageing in dense colloids as diffusion in the logarithm of time, J. Phys. Condens. Matter, 23, 6, 065103 (2011)
\bibitem{bib:doliwa} B. Doliwa, A. Heuer, Cooperativity and spatial correlations near the glass transition: Computer simulation results for hard spheres and disks, Phys. Rev. E, 61, 6, 6898 (2000)
\bibitem{bib:weeks} E. R. Weeks, J. C. Crocker, D. A. Weitz, Short-and long-range correlated motion observed in colloidal glasses and liquids, J. Phys.: Condens. Matter, 19, 20, 205131 (2007)
\bibitem{bib:dim} S. Vivek, C. P. Kelleher, P. M. Chaikin, E. R. Weeksa, Long-wavelength fluctuations and the glass transition in two dimensions and three dimensions, PNAS, 114, 8, 1850 (2017)
\bibitem{bib:cluster} C. Donati, S. C. Glotzer, P. H. Poole, W. Kob, S. J. Plimpton, Spatial correlations of mobility and immobility in a glass-forming Lennard-Jones liquid, Phys. Rev. E, 60, 3, 3107 (1999)
\bibitem{bib:coop} C. Dalle-Ferrier, C. Thibierge, C. Alba-Simionesco, L. Berthier, G. Biroli, J. P. Bouchaud, F. Ladieu, D. L'H\^{o}te, G. Tarjus, Spatial correlations in the dynamics of glassforming liquids: Experimental determination of their temperature dependence, Phys. Rev. E, 76, 4, 041510 (2007)
\bibitem{bib:donth} E. Donth, H. Huth, M. Beiner, Characteristic length of the glass transition, J. Phys.: Condens. Matter, 13, 22, L415 (2001)
\bibitem{bib:nature} W. Kob, S. Rold\'an-Vargas, L. Berthier, Non-monotonic temperature evolution of dynamic correlations in glass-forming liquids, Nature Physics, 8, 164 (2012)
\bibitem{bib:mct} D. R. Reichman, P. Charbonneau, Mode-coupling theory, J. Stat. Mech., 2005, 05, P05013 (2005)
\bibitem{bib:rfot} T.R. Kirkpatric, D. Thirumalai, Colloquium: Random first order transition theory concepts in biology and physics, Rev. Mod. Phys., 87, 1, 183, (2015)
\bibitem{bib:4point}  L. Berthier, G. Biroli, J.-P. Bouchaud, L. Cipelletti, D. El Masri, D. L'H\^{o}te, F. Ladieu, M. Pierno, Direct experimental evidence of a growing length scale accompanying the glass transition, Science, 310, 5755, 1797 (2005)
\bibitem{bib:relax} P. Luo, P. Wen, H.Y. Bai, B. Ruta, W.H. Wang, Relaxation decoupling in metallic glasses at low temperatures, Phys. Rev. Lett., 118, 22, 225901 (2017)
\bibitem{bib:aging} Y. Lahini, O. Gottesman, A. Amir, S. M. Rubinstein, Nonmonotonic aging and memory retention in disordered mechanical systems, Phys. Rev. Lett. 118, 8, 085501 (2017)
\bibitem{bib:spin} G. Parisi, The order parameter for spin glasses: a function on the interval 0-1, J. Phys. A, 13, 3, 1101 (1980)
\bibitem{bib:leutheusser} E. Leutheusser, Dynamical model of the liquid-glass transition, Phys. Rev. A 29, 5, 2765 (1984)
\bibitem{bib:mosayebi} M. Mosayebi, E. Del Gado, P. Ilg, H. C. \"Ottinger, Probing a critical length scale at the glass transition, Phys. Rev. Lett., 104, 20, 205704 (2010)
\bibitem{bib:wyart} M. Wyart, M. E. Cates, Does a growing static length scale control the glass transition?, Phys. Rev. Lett., 119, 19, 195501 

\bibitem{bib:majka2} M. Majka, P.F. G\'{o}ra, Non-Gaussian polymers described by alpha-stable chain statistics: Model, effective interactions in binary mixtures, and application to on-surface separation, Phys. Rev. E 91, 5, 052602 (2015)
\bibitem{bib:binmix_cryst} S. Martin, G. Bryant, and W. van Megen, Crystallization kinetics of polydisperse colloidal hard spheres. II. Binary mixtures, Phys. Rev. E, 71, 2, 021404 (2005)
\bibitem{bib:binmix2} J. Hendricks, R. Capellmann, A. B. Schofield, S. U. Egelhaaf, and M. Laurati, Different mechanisms for dynamical arrest in largely asymmetric binary mixtures, Phys. Rev. E 91, 3, 032308 (2015)
\bibitem{bib:binmix1} R. Ju\'{a}rez-Maldonado and M. Medina-Noyola, Theory of dynamic arrest in colloidal mixtures, Phys. Rev. E 77, 5, 051503 (2008) 
\bibitem{bib:binmix3} S. R. Williams and W. van Megen, Motions in binary mixtures of hard colloidal spheres: Melting of the glass, Phys. Rev. E 64, 4, 041502 (2001)
\bibitem{bib:binmix2D} D. N. Perera and P. Harrowell, Stability and structure of a supercooled liquid mixture in two dimensions, Phys. Rev. E 59, 5, 5721 (1999)
\bibitem{bib:binmix_c} R. Benzi, M. Sbragaglia, M. Bernaschi, and S. Succi, Phase-Field Model of Long-Time Glasslike Relaxation in Binary Fluid Mixtures, Phys. Rev. Lett. 106, 16, 164501 (2011)
\bibitem{bib:binmix_q}  B.Gadway, D. Pertot, J. Reeves, M. Vogt, and D. Schneble, Glassy Behavior in a Binary Atomic Mixture, Phys. Rev. Lett. 107, 14, 145306 (2011)

\bibitem{bib:kubo} R. Kubo, M. Toda, N. Hashitsume, The fluctuation-dissipation theorem in \emph{Statistical Physics II Nonequilibrium Statistical Mechanics}, 37, ed. H. K. V. Lotsch (Springer-Verlag, 1991)

\bibitem{bib:inertial_to_overdamped} X. Durang, C. Kwon, and H. Park, Overdamped limit and inverse-friction expansion for Brownian motion in an inhomogeneous medium, Phys. Rev. E 91, 6, 062118 (2015)
\bibitem{bib:gardiner} C. W. Gardiner, \emph{Handbook of stochastic methods} ed. H. Haken, (Springer-Verlag, 2004)
\bibitem{bib:lau} A.W.C. Lau, T.C. Lubensky, State-dependent diffusion: Thermodynamic consistency and its path integral formulation, Phys. Rev. E, 76, 1, 011123 (2007)
\bibitem{bib:oded} O. Farago, N. Gr{\o}nbech-Jensen, Fluctuation-Dissipation Relation for Systems with Spatially Varying Friction, J. Stat. Phys., 156, 6, 1093 (2014)

\bibitem{bib:mct_bad} L. Berthier and G. Tarjus, Critical test of the mode-coupling theory of the glass transition, Phys. Rev. E, 82,3, 031502 (2010)
\bibitem{bib:szamel} G. Szamel, Theory for the dynamics of dense systems of athermal self-propelled particles, Phys. Rev. E, 93,1, 012603 (2016)
\bibitem{bib:feng} M. Feng, Z. Hou, Mode coupling theory for nonequilibrium glassy dynamics of thermal self-propelled particles, Soft Matter, 13, 25, 4464 (2017)
\bibitem{bib:exact_mct} R. van Zon and J. Schofield, Phys. Rev. E, 65, 1, 011106 (2001)
\bibitem{bib:yodh} A. G. Yodh, K. Lin, J. C. Crocker, A. D. Dinsmore, R. Verma, P. D. Kaplan, Entropically driven self-assembly and interaction in suspension, Phil. Trans. R. Soc. Lond. A 359, 1782, 921 (2001)
\bibitem{bib:grier} D. G. Grier, A revolution in optical manipulation, Nature (London) 424, 6950, 810 (2003)
\bibitem{bib:ciesla} M. Cie\'sla, J. Barbasz, Random packing of spheres in Menger sponge, J. Chem. Phys., 138, 10, 214704 (2013)
\bibitem{bib:rcp} S. Torquato, T. M. Truskett, and P. G. Debenedetti, Is random close packing of spheres well defined?, Phys. Rev. Lett. 84, 10, 2064 (2000)
\bibitem{bib:jamming} A. Ikeda, L. Berthier and P. Sollich, Unified study of glass and jamming rheology in soft particle systems, Phys. Rev. Lett., 109,1, 018301 (2012)
\bibitem{bib:2d} F. Weysser and D. Hajnal, Tests of mode-coupling theory in two dimensions, Phys. Rev. E 83, 4, 041503 (2011)
\bibitem{bib:2d_up} S. Meyer, C. Song, Y. Jin, K. Wang, H. A.Makse, Jamming in two-dimensional packings, Physica A, 389, 22, 5137 (2010)
\bibitem{bib:length} K. H. Nagamanasa, S. Gokhale, A. K. Sood, R. Ganapathy, Direct measurements of growing amorphous order and non-monotonic dynamic correlations in a colloidal glass-former, Nat. Phys., 11, 5, 403 (2017)




\end{thebibliography}
\end{document}